\PassOptionsToPackage{dvipsnames}{xcolor}
\documentclass[twocolumn,twocolappendix]{aastex631}

\usepackage{graphicx}
\usepackage[dvipsnames]{xcolor}
\usepackage{todonotes}
\usepackage{multirow}
\usepackage{amsmath}
\usepackage{amssymb}
\usepackage{mathrsfs}

\usepackage{siunitx}
\usepackage{hyperref}
\usepackage{soul}
\DeclareSIUnit\angstrom{\text {Å}}

\newcommand{\kms}{\ensuremath{\mathrm{km~s}^{-1}}}
\newcommand{\mbh}{\ensuremath{M_\mathrm{BH}}}
\newcommand{\mlgen}{\ensuremath{M^*/L}}
\newcommand{\msun}{\ensuremath{M_{\odot}}}
\newcommand{\mgb}{Mg~{\sc i}~$b$}

\newcommand{\Hae}{H15A}

\newcommand{\imfit}{\textsc{Imfit}}
\newcommand{\tmin}{T_{\mathrm{min}}}
\newcommand{\tmaj}{T_{\mathrm{maj}}}

\newcommand{\sersic}{S{\'e}rsic}
\newcommand{\TriOS}{TriOS}

\newcommand{\mdm}{\ensuremath{M_{\mathrm{50}}}}

\begin{document}

\title{
A 22-Billion Solar Mass Black Hole in Holmberg 15A with Keck KCWI Spectroscopy and Triaxial Orbit Modeling
}

\correspondingauthor{Emily Liepold}\email{emilyliepold@berkeley.edu}

\author
[0000-0002-7703-7077]
{Emily R. Liepold}
\affiliation{Department of Astronomy, University of California, Berkeley, California 94720, USA.}

\author
[0000-0002-4430-102X]
{Chung-Pei Ma}
\affiliation{Department of Astronomy, University of California, Berkeley, California 94720, USA.}
\affiliation{Department of Physics, University of California, Berkeley, California 94720, USA.}

\author
[0000-0002-1881-5908]
{Jonelle L. Walsh}
\affiliation{George P. and Cynthia Woods Mitchell Institute for Fundamental Physics and Astronomy, and Department of Physics and Astronomy, Texas A\&M University, College Station, TX 77843, USA.}

\begin{abstract}

Holmberg 15A (\Hae), the brightest cluster galaxy of Abell~85, has an exceptionally low central surface brightness even among local massive elliptical galaxies with distinct stellar cores, making it exceedingly challenging to obtain high-quality spectroscopy to detect a supermassive black hole (SMBH) at its center.
Aided by the superb sensitivity and efficiency of KCWI at the Keck II Telescope, we have obtained spatially resolved stellar kinematics over a ${\sim}100''\times 100''$ contiguous field of H15A for this purpose.
The velocity field exhibits a low amplitude (${\sim}20$ \kms) rotation along a kinematic axis that is prominently misaligned from the photometric major axis, a strong indicator that \Hae\ is triaxially shaped with unequal lengths for the three principal axes.
Using 2500 observed kinematic constraints, we perform extensive calculations of stellar orbits with the triaxial Schwarzschild code, TriOS, and search over ${\sim}$40,000 galaxy models to simultaneously determine the mass and intrinsic 3D shape parameters of \Hae.  
We determine a ratio of $p=0.89$ for the middle-to-long principal axes and $q=0.65$ for the short-to-long principal axes. 
Our best estimate of the SMBH mass, $\mbh=(2.16^{+0.23}_{-0.18})\times 10^{10}\msun$, makes H15A -- along with NGC~4889 -- the galaxies hosting the most massive SMBHs known in the local universe.
Both SMBHs lie significantly above the mean $\mbh-\sigma$ scaling relation.
Repeating the orbit modeling with the axisymmetrized version of TriOS produces worse fits to the KCWI kinematics and increases \mbh\ to $(2.55\pm 0.20) \times 10^{10}\msun$, which is still significantly below $\mbh=(4.0\pm 0.8) \times 10^{10}\msun$ reported in a prior axisymmetric study of \Hae.

\end{abstract}


\section{Introduction} \label{sec:intro}

The central light profiles of stars in massive elliptical galaxies have flattened cores that differ significantly from the cuspy centers of less massive elliptical galaxies and bulges of disk galaxies. These stellar cores are a defining feature of the most massive elliptical galaxies (e.g., \citealt{faberetal1997, laueretal1995, laueretal2005}), indicating a significant deficit of stars relative to a standard \sersic\ profile. A possible mechanism for flattening the inner stellar distribution is three-body gravitational scattering that repeatedly slingshots stars passing close to a black hole binary to larger radii (e.g., \citealt{begelmanetal1980, kormendybender2009, thomasetal2014}).

The stellar core size of a massive elliptical galaxy is correlated with the mass of its central supermassive black hole (SMBH), \mbh, and is a more robust estimator of \mbh\ for these galaxies than the stellar velocity dispersion, $\sigma$ (e.g., \citealt{laueretal2007, ruslietal2013, thomasetal2016}). The $\mbh -\sigma$ relation followed by lower-mass SMBHs in bulge-dominated galaxies \citep{gebhardtetal2000a, ferraresemerritt2000} exhibits large scatter at $\sigma \ga 270$ \kms\ \citep[e.g.,][]{laueretal2007, mcconnellma2013, kormendybender2009}, suggesting different evolutionary paths for the most massive galaxies and their central SMBHs.  The stellar core size, on the other hand, correlates with \mbh\ to the highest masses.  Furthermore, the stellar core radius is also found to be tightly correlated with the radius of the SMBH's gravitational sphere of influence (SOI), $r_{\rm SOI}$, the location at which the enclosed stellar mass equals \mbh: the best-fit relation is consistent with the two radii being equal, with a small intrinsic scatter of 0.17 dex \citep{thomasetal2016}.  

In this paper, we study Holmberg 15A (\Hae), a massive elliptical galaxy first cataloged in \citet{Holmberg1937}. \Hae\ is the brightest cluster galaxy (BCG) at the center of Abell~85 at a redshift of $z = 0.0555$. It is in the collection of 116 BCGs of nearby Abell clusters in the two-color photometric study of \citet{hoessel_paper_I_1980}. When the surface brightness profiles of these galaxies are parameterized by a modified Hubble form, $I(R)=I_c/(1+R^2/R_c^2)$, \Hae\ is found to have the largest physical core radius of $R_c=6.2$ kpc (corrected to our adopted $H_0$; \citealt{Hoessel1980}). \Hae\ is therefore a potential host for an enormous SMBH and has been reported to harbor one with $\mbh=(4.0\pm 0.8) \times 10^{10}\msun$ \citep{mehrganetal2019}.

Subsequent studies, however, have reported conflicting core sizes when a variety of functional forms is used to approximate the surface brightness profile of \Hae. \citet{lopez-cruzetal2014} fit the Nuker form \citep{laueretal1995} to $R$-band photometry from the Kitt Peak National Observatory and $r$-band photometry from the Canada-France-Hawaii Telescope (CFHT). For both data sets, they find the core size to be well parameterized by a radius of $r_\gamma = 4.57$ kpc (at which the local logarithmic slope of the surface brightness is $-1/2$) and suggest \Hae\ is the host of a candidate SMBH with mass as large as $\sim 3\times 10^{11}\msun$. \citet{Bonfini2015}, on the other hand, find the same CFHT light profile to be well fit by a \sersic\ profile of a low index $n$ plus an exponential outer component without the need for a core parameter. Using Gemini Multi Object Spectrograph (GMOS) imaging data in better seeing conditions, \citet{madriddonzelli2016} find a similar value of $r_\gamma=5.02$ kpc as \citet{lopez-cruzetal2014} for the Nuker form, but a double \sersic\ form provides a better overall fit. \citet{mehrganetal2019} find $r_\gamma =4.11 \pm 0.11$ kpc from their photometric data. When they instead fit a three-component model including a central core-\sersic\ component, this component has an index of $n \sim 1$ and a break radius of $r_b = 2.8$ kpc.  

The core size debate in the literature aside, another indication of the presence of an ultra-massive SMBH at the center of \Hae\ comes from its exceptionally low central surface brightness and stellar density. The surface brightness of \Hae\ (see Section~3) varies slowly in the central few arcseconds, with $\mu_V \approx 20.1$ mag arcsec$^{-2}$ between a radius of $0.5''$ and $2''$. In comparison, the host galaxies of two ultra-massive SMBHs in the local universe, both with $\mbh = (1.7-2)\times 10^{10} \msun$ from stellar dynamical measurements, have a central surface brightness of $\mu_V = 17.6$ mag arcsec$^{-2}$ (NGC~4889; \citealt{mcconnelletal2011a}) and $\mu_V = 17.8$ mag arcsec$^{-2}$ (NGC~1600; \citealt{thomasetal2016}). In physical units, the stellar surface density is $\mu= (1500-2000)\,L_\sun\, {\rm pc}^{-2}$ at a radius of 1 kpc for NGC~1600 and NGC~4889, while it is $\mu = 320 \,L_\sun\, {\rm pc}^{-2}$ for \Hae.

The low central surface brightness of \Hae\ makes it extremely challenging to obtain the high signal-to-noise ratio ($S/N$) spectroscopic data required to study its central SMBH.  To this end, we have conducted extensive spectroscopic observations of \Hae\ using the Keck Cosmic Web Imager (KCWI) on the Keck II Telescope in five observational runs from 2018 to 2021. The exquisite sensitivity of this instrument enables us to co-added spectra for 313 contiguous spatial bins over a ${\sim}100''\times 100''$ field of view (FOV) of \Hae. The central $8.25''\times 20''$ FOV is covered finely by 97 spatial bins with $S/N$ above 130; the rest of the FOV is covered by 216 bins with $S/N\sim 60$.  The full shape of the line-of-sight stellar velocity distribution (LOSVD) is measured for each spectrum. Using the resulting ${\sim}2500$ kinematic moments as constraints, we perform  triaxial Schwarzschild orbit modeling with the TriOS code \citep{quennevilleetal2022} to simultaneously determine the black hole mass, stellar mass-to-light ratio, dark matter halo mass, and the 3D intrinsic shape of \Hae.

The Keck observations, KCWI data reduction, and stellar kinematic measurements are described in Section~\ref{sec:spectroscopic_observations}. Photometric properties of \Hae\ based on archival data are presented in Section~\ref{sec:photometry}. We summarize the triaxial orbit modeling method and galaxy parameter search algorithm in Section~\ref{sec:triaxial_orbit_models}, present the best-fit galaxy parameters in Section~\ref{sec:H15_results}, and discuss the scaling relations of \mbh\ and host galaxy properties in Section~\ref{sec:scaling}. In Section~\ref{sec:discussion}, we perform detailed comparisons with the prior dynamical study of \Hae\ SMBH using Multi Unit Spectroscopic Explorer (MUSE) on the Very Large Telescope \citep{mehrganetal2019}.

\Hae's redshift $z = 0.0555$ corresponds to an angular diameter distance of 222.3 Mpc for a flat $\Lambda$CDM cosmology with $H_0 = 70\, \kms \text{ Mpc}^{-1}$ and $\Omega_m = 0.31$. An angular size of $1''$ spans a physical size of 1.078 kpc at this distance.

\section{Stellar Kinematics from Keck KCWI Observations} \label{sec:spectroscopic_observations}

\subsection{KCWI Observations} \label{sec:observations}

We observed \Hae\ using the Keck Cosmic Web Imager (KCWI; \citealt{morrisseyetal2018}) on the Keck II Telescope in November 2018, November 2019, November 2020, September 2021, and November 2021. We used the BL grating of the integral field spectrograph with a central wavelength of 4600 \AA\ to achieve the widest spectral coverage, about 3600\text{--}5600 \AA\ (3410\text{--}5305 \AA\ in \Hae's rest frame). This spectral region contains numerous stellar absorption features including Ca H and K, the G band, H$\beta$, and the \mgb\ triplet region.

During the November 2018 run, we used the small slicer of KCWI to observe the central region of \Hae. This configuration has a FOV of $8\farcs25\times 20''$ with $0\farcs34\times 0\farcs30$ spatial pixels. Seven exposures totaling $\SI{150}{\minute}$ were taken at slightly dithered positions. Three sky exposures totaling $\SI{30}{\minute}$ were also taken following an object-sky-object pattern. The slicer was oriented with the $20''$ axis lying along the photometric major axis at a position angle (PA) of $-34^\circ$. The atmospheric seeing was 0\farcs60, as measured by the differential image motion monitor at the nearby CFHT.

\begin{figure*}
\centering
\includegraphics[width=1.0\linewidth]{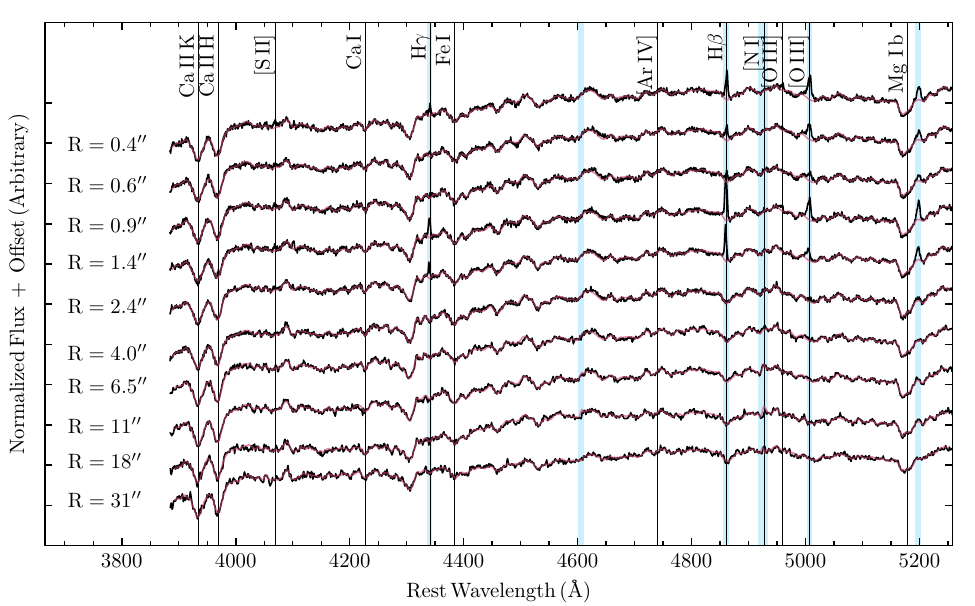}
\caption{
Ten representative sky-subtracted KCWI spectra (black curves) of \Hae\ from spatial bins at increasing distance from the galaxy's center ($0.4''$ to $31''$ from top to bottom).
The inner six spectra are from the KCWI small slicer; the outer four are from the large slicer.
Each spectrum is obtained from co-adding spectra from individual KCWI spaxels to meet a $S/N$ threshold.
Our observations provide a total of 313 co-added spectra for stellar orbit modeling:
97 spectra with $S/N \ga 130$ per 0.5 \AA\ from the small slicer,
and 216 spectra with $S/N \ga 60$ per \AA\ from the large slicer. 
For each spectrum, the stellar templates broadened by the best-fit LOSVD (red curves) provide a close match.
Four emission lines and two sky lines (vertical blue bands) are masked out and excluded from the spectral fitting.
}
\label{fig:spectra}
\end{figure*}  

During the other four observing runs, we used the large slicer of KCWI to achieve a wide spatial coverage.  The FOV is four times larger in this configuration, covering $33'' \times 20''$, with $1\farcs38\times 0\farcs30$ spatial pixels. We used a mosaic of 29 exposures with 10 pointings to cover ${\sim}100''$ (${\sim}108$ kpc) along both the photometric major and minor axes and ${\sim}85''$ (${\sim}92$ kpc) midway between those axes. These total exposure time is $\SI{574}{\minute}$. An additional 18 sky exposures totaling $\SI{180}{\minute}$ were taken. The seeing measurements for the large slicer observations were similar, between 0\farcs57 and 0\farcs70.

Only data taken in good observing conditions are used in the following analysis. The total exposure times for the science and sky frames of these data are 12.1 hr and 3.5 hr, respectively. 

\subsection{KCWI Data Reduction} \label{sec:reduction}

We perform the initial round of data processing using the KCWI Data Extraction and Reduction Pipeline (KCWI-DRP; \citealt{morrisseyetal2018}). This pipeline performs overscan and bias removal, cosmic ray rejection, dark and scattered light subtraction, finding the geometric distortion and wavelength solution, flat-fielding, correction for vignetting and the illumination pattern, sky subtraction, and the generation of three-dimensional data cubes. After these steps, the cubes are corrected for differential atmospheric correction and flux calibrated by use of a standard star. 

We perform a series of additional customized reduction steps to properly handle sky subtraction and science frame mosaics. Similar procedures are used and described in our KCWI study of M87 \citep{liepoldetal2023}. The exposures are corrected for a small number of cosmic rays that are improperly removed by the KCWI DRP. We then construct a series of representative sky spectra by performing a principal component analysis (PCA) on all available sky spectra. These PCA components are included as additional additive components in the spectral fitting in Section~\ref{sec:kinematics} to correct for residual contributions from the sky that persist after the coarse sky subtraction performed in KCWI-DRP.  Finally, the science frames are mosaicked onto a common $0\farcs15\times0\farcs17\times0.5$ \AA\ grid for the small slicer exposures and a common $0\farcs3\times1\farcs4\times1$ \AA\ grid for the large slicer exposures. 

Both spatial and spectral masking are applied to the data cube. Several regions within the spatial mosaic are contaminated by compact foreground objects. To mask these regions, we collapse the data cubes spectrally, locate regions with substantially higher flux than their surroundings, and then iteratively mask the brightest pixels in the region until the contaminant is fully removed. The emission lines present in the central ${\sim}3''$ are masked out: H$\gamma$ (4340 \AA), H$\beta$ (4861 \AA), [O III] (5007 \AA), and [N I] (5200 \AA). Two highly variable sky lines are also masked out: 4861 \AA~H$\beta$ (4605 \AA\ in \Hae's rest frame) and 5200 \AA~[N~I] (4926 \AA\ rest).  

We determine the line-spread function (LSF) using FeAr arc lamp exposures obtained during instrument calibration before each night of observations. For each slicer, we perform simultaneous fits to the arc lamp spectra around 31 known spectral lines between 4500 \AA\ and 5000 \AA. The small slicer's LSF is well approximated as a Gaussian with a FWHM of 1.35 \AA\ in our spectral window. We find the large slicer's LSF to be distinctly non-Gaussian; instead, it is well characterized as the convolution of a top-hat function with a half-width of 2.55 \AA\ and a Gaussian function with FWHM $1.46$ \AA\ (see Appendix~A of \citealt{liepoldetal2023}).

\subsection{Stellar Spectra and Kinematics} \label{sec:kinematics}

\begin{figure*}
\centering
\includegraphics[width=0.9\textwidth]{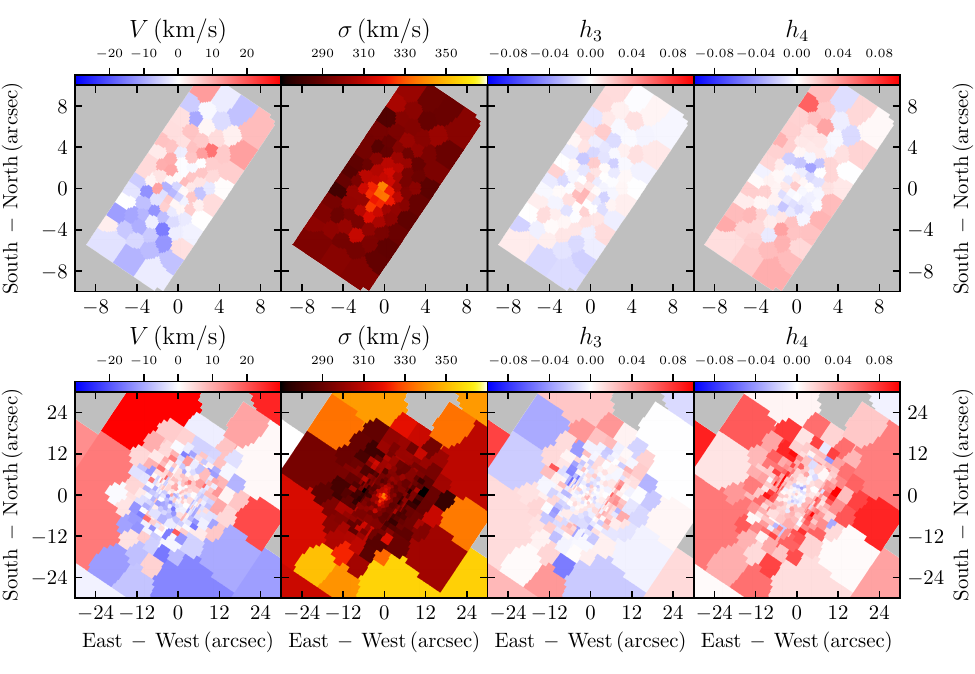}
\caption{
Maps of the first four Gauss-Hermite moments, $V, \sigma, h_3$, and $h_4$ (left to right), of the stellar LOSVDs of \Hae\ as measured for 313 spatial bins from the Keck KCWI spectra.
The top row displays the zoomed-in central $20''\times8''$ region covered by the KCWI small slicer;
the bottom row displays the $60''\times 60''$ region from the small and large slicer mosaics.
North is up and east is left.
}
\label{fig:H15_GH_moment_maps}
\end{figure*}

We measure the LOSVDs from the shapes of the absorption features in the KCWI spectra. To obtain spectra of uniformly high $S/N$ for this measurement, we construct a set of spatial bins for each slicer using the Voronoi binning scheme \citep{cappellaricopin2003} and co-add the spectra from individual spaxels within each bin. For the small slicer data, the target $S/N$ is ${\sim} 130$ per 0.5 \AA\ spectral pixel, resulting in 97 spatial bins. For the large slicer, a target of $S/N \sim 60$ per 1.0 \AA\ spectral pixel results in 216 bins outside the spatial region covered by the small slicer observations.  These target $S/N$ values are chosen so that the innermost several bins are comparable in size to the seeing conditions, and the areas of individual bins increase relatively smoothly from the central region of the galaxy covered by the small slicer to the outer region covered by the large slicer. Ten representative spectra at increasing distances from the galaxy center are displayed in Figure~\ref{fig:spectra}.

\begin{figure}
\centering
\includegraphics[width=\linewidth]{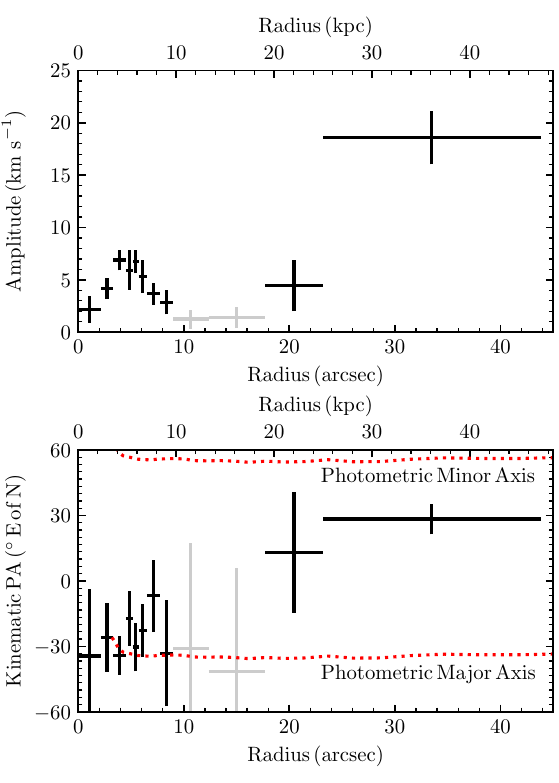}
\caption{
Misalignment between the kinematic and photometric axes of \Hae.
We model the KCWI velocity field with $V(R,\Theta) = V_1(R) \cos \left[ \Theta - \Theta_0(R) \right]$ and measure the amplitude $V_1(R)$ (upper panel) and phase $\Theta_0(R)$ (lower panel) of the rotation pattern.
The phase $\Theta_0$ determines the PA of the kinematic axis, ${\rm PA}_{\rm kin}$, which varies with increasing radius and reaches ${\rm PA}_{\rm kin}= 28^\circ$ beyond $20''$, leading to a misalignment of $62^\circ$ from the photometric major and minor axes (red dotted lines).
Between $9''$ and $18''$, the measured amplitude is consistent with 0: $V_1 = 1.32 \pm 0.95$ \kms (grey).
}
\label{fig:H15_vtheta_fit}
\end{figure}

We use pPXF \citep{Cappellari2017} to determine the LOSVD from the co-added spectrum for each spatial bin. The LOSVD is characterized as a Gauss-Hermite series, and the coefficients of the first eight terms in the series, $V, \sigma, h_3, \dots$ and $h_8$, are parameters in the fit. A multiplicative polynomial of order 7 is used to model the continuum shape of each spectrum. The first ten most significant PCA sky components are used as additional additive terms whose weights are chosen during fitting. For the stellar templates, we use a set of 671 ``clean" Indo-US stellar spectra \citep{valdesetal2004} that have well determined spectral types listed in their Table~2. The spectral resolution of the Indo-US stellar templates is 1.35 \AA\ FWHM, which happens to be the same as that of the KCWI small slicer, so no adjustment is needed during spectral fitting of the small slicer data. For the KCWI large slicer data, each stellar template spectrum is first broadened to match the large slicer's LSF (see Section~\ref{sec:reduction}) and then used to measure the LOSVDs.

Maps of the first four moments of the LOSVDs of \Hae\ are presented in Figure~\ref{fig:H15_GH_moment_maps}. The map of the lowest moment, the line-of-sight velocity, shows coherent red and blue sides with a velocity difference of ${\sim}20$ \kms.  To further quantify the kinematic axis and amplitude, we model the velocity field with $V(R,\Theta) = V_1(R) \cos \left[ \Theta - \Theta_0(R) \right]$ and determine the velocity amplitude, $V_1(R)$, and the PA of the kinematic axis, $\Theta_0(R)$, using groups of bins at different distances $R$ from the center of the galaxy \citep{liepoldetal2023}. The resulting profiles are plotted in Figure~\ref{fig:H15_vtheta_fit}. Within $R \sim 10''$, there is a very low amplitude rotation of up to $V \sim 7$ \kms; the PA of the kinematic axis (defined as the PA where the velocity is maximized) in this central region is consistent with the photometric major axis with $\rm{PA}_{\rm photo}=-34^\circ$. Between $R\sim 9''$ and $18''$, no coherent rotation is detected in our data, hence the undetermined kinematic PA. From $R \sim 20''$ to the outer reach of our data at $R \sim 45''$, the velocity amplitude rises to ${\sim}20$ \kms\ and the kinematic PA exhibits a significant twist, reaching $\rm{PA}_{\rm kin}=28^\circ \pm 7^\circ$ and resulting in a large kinematic misalignment from the photometric major axis: $\Psi \equiv |\rm{PA}_{\rm kin}-\rm{PA}_{\rm photo}| \approx 62^\circ$. We have performed alternative fits to the velocity map using \texttt{kinemetry} \citep{krajnovicetal2006} and found similar behavior in the amplitude and phase profiles.

Axisymmetric models by construction produce aligned kinematic and photometric axes and therefore cannot account for the observed kinematic misalignment of H15. Kinematic misalignment is a common feature in massive early-type galaxies. In a study of the stellar kinematics of 90 galaxies in the MASSIVE survey \citep{maetal2014}, \citet{eneetal2018} find 76\% of the sample to be slow- or non-rotators with a spin parameter $\lambda < 0.2$. Among the slow rotators with measurable kinematic axes, 57\% have misaligned kinematic axes nearly evenly distributed between $\Psi=15^\circ$ and $90^\circ$. The presence of a significant kinematic misalignment in the slowly-rotating \Hae\ is thus consistent with expectations. 

The radial profiles of the eight kinematic moments are presented in Figure~\ref{fig:H15_kinematic_profile} of Appendix~\ref{appendix:kinematic_profiles}.  A detailed comparison of the velocity dispersion measurements with those found by \citet{mehrganetal2019} is given in Section~\ref{sec:stellar_kinematic_comparison}. The central $8\farcs25\times 20''$ of the innermost science frames observed with the large slicer overlap with the FOV of the small slicer. Since the smaller slicer data have better angular and spatial resolution, we do not include the (redundant) large-slicer data in this overlapping region in the following analysis. We have examined that the stellar kinematics from the two slicers in this common region are consistent with each other.

\section{Photometric Properties} \label{sec:photometry}

To characterize the stellar light distribution of \Hae, we use archival observations from GMOS on the Gemini South telescope (PI Madrid;  \citealt{madriddonzelli2016}). The instrument was operating in its imaging mode with the filter \emph{r\_G0326} (similar to SDSS $r'$, and referred to as $r'$ below). Two exposures of $\SI{200}{\second}$ were obtained with an effective pixel scale of $0\farcs 16$  per pixel and a seeing of $0\farcs 56$. We use data from only the central GMOS CCD chip to ensure that the background sky is uniform. This results in a $282''\times 159''$ field, centered on \Hae\ with the long axis of the field aligned with north. We perform spatial masking to exclude contaminants and companion objects. The companion galaxy located at $75''$ to the northwest of the center of \Hae\ has been modeled and subtracted from the image before performing the primary fitting of \Hae. 

\begin{figure}[htbp]
\centering
\includegraphics[width=0.8\linewidth]{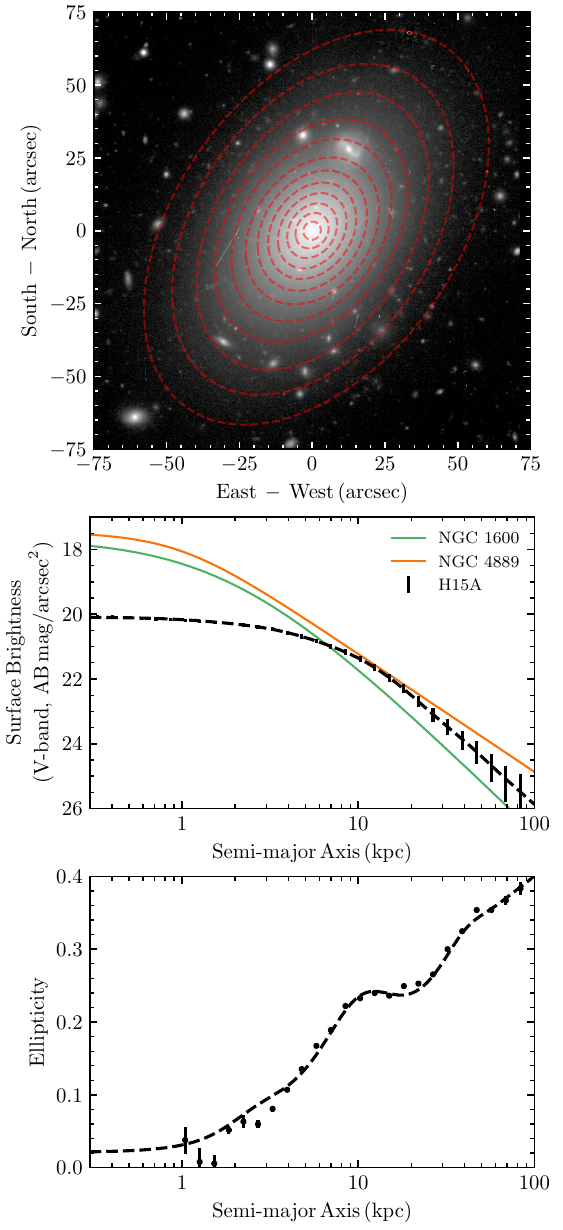}
\caption{
Photometry of \Hae.
(Top) Gemini image of the central $150''\times 150''$ region and representative isophotes (dashed red curves).
(Middle) Surface brightness profile along the semi-major axis of \Hae. The MGE model fit to the data is marked with a dashed curve.
A color of $V-r'=0.5$ is assumed to convert the observed $r'$-band to $V$-band.  For comparison, two cored massive elliptical galaxies known to harbor ultra-massive SMBHs are shown: NGC~1600 (green) and NGC~4889 (orange).
The central surface brightness of \Hae\ is more than 2 magnitudes fainter.
(Bottom) The isophotes of \Hae\ are nearly round within the central few arcseconds but flatten significantly to $\epsilon \sim 0.4$ at $100''$. The dashed curve denotes the ellipticity of the MGE model.
}
\label{Fig:H15_photometry}
\end{figure}

We perform isophotal fits \citep{jedrzejewski1987} using the \texttt{ellipse} algorithm implemented in the photutils Python package \citep{photutils} to measure the photometric properties of \Hae.  As Figure~\ref{Fig:H15_photometry} shows, the isophotes are nearly round at the center but flatten significantly with radius, where the ellipticity increases from $\epsilon \la 0.05$ within $1''$ to $\epsilon = 0.38$ at $100''$ (bottom panel). In the central region, the round isophotes lead to large uncertainty in the photometric PA. Beyond $3''$, the photometric PA is essentially constant with ${\rm PA_{photo}} = -34^\circ \pm 2^\circ$ to the outermost isophotes at $100''$ (red dotted curve in lower panel of Figure~\ref{fig:H15_vtheta_fit}).

The luminosity density of \Hae\ is used to construct the stellar gravitational potential in orbit modeling in Section~\ref{sec:triaxial_orbit_models}. It is obtained from deprojection of the surface brightness parameterized by a Multi-Gaussian expansion (MGE). Details of the MGE determination, including corrections made to remove unphysically large central luminosity densities and unphysically flat outer ellipticity, are discussed in Appendix~\ref{appendix:mge}. The MGE fits to the surface brightness and ellipticity profiles are plotted as dashed curves in the lower two panels of Figure~\ref{Fig:H15_photometry}. 

Integration of the MGE fit yields a total luminosity of $L_{*,r'} =6.0\times10^{11}L_{\odot}$ in $r'$-band, or $L_{*,\emph{V}} =3.8\times10^{11}L_{\odot}$ in $V$-band, assuming $V-r'=0.5$ for a ${\sim}13$ Gyr stellar population \citep{vazdekisetal2016, edwardsetal2016}. We determine the half-light (effective) radius of \Hae\ to be $R_e = 35\farcs4 = 38.1$ kpc.

\section{Mass and Shape Determinations} \label{sec:orbit}

\subsection{Triaxial Orbit Models} \label{sec:triaxial_orbit_models}

We use the \TriOS\ code \citep{quennevilleetal2022} to construct triaxial Schwarzschild orbit superposition models of \Hae.  Our recent study based on 25 sets of simulated galaxy kinematics indicates that this code -- when combined with our parameter sampling and inference schemes -- can robustly recover the input galaxy mass and shape parameters \citep{pilawaetal2024}.

A galaxy model in \TriOS\ consists of multiple mass components including a central black hole of mass \mbh, a stellar component described by the deprojected surface brightness of \Hae\ and a stellar mass-to-light ratio \mlgen, and a dark matter halo. We choose to parameterize the dark matter halo with a generalized mass density profile of \cite{Navarroetal1996}: 
$\rho(r) = \rho_0 (r / r_s)^{-\gamma} (1 + r/r_s)^{\gamma - 3}$,
where $r_s$ is the scale radius and $\gamma$ is the central logarithmic slope. We set $\gamma = 0$ so that the dark matter halo has a finite central density, $\rho_0$, and a flattened density profile in the inner region in resemblance of \Hae's stellar distribution. Our data are more sensitive to the total enclosed dark matter mass than individual halo parameters, so we fix $r_s = 150$ kpc and let $\rho_0$ be a free parameter (see discussion in Section~\ref{sec:H15_results}). For ease of interpretation, we report the halo properties in terms of its mass enclosed within 50 kpc (the outer reach of the KCWI data) denoted by \mdm, which is linearly proportional to $\rho_0$.

In addition to the mass parameters, a triaxial galaxy model requires three angles, $\theta, \phi$ and $\psi$, to relate the intrinsic and projected coordinate systems and specify the deprojection of the light distribution. Alternatively, three dimensionless shape parameters, $T, \tmaj$ and $\tmin$, can be used, which are related to the angles by Equation~(8) in \citet{quennevilleetal2022}. This triplet of shape parameters is convenient to use during the parameter search as it maps all valid deprojections into a unit cube and reduces the degeneracy among the three parameters. The deprojection can also be specified by a set of three axis ratios: ratio of the middle-to-long principal axis lengths, $p = b / a$, ratio of the short-to-long principal axis lengths, $q = c / a$, and ratio of projected and intrinsic axis lengths, $u = a' / a$, which represents a compression scale factor. Each deprojected component of the MGE is allowed its own $p, q$ and $u$ in order to account for the radially varying ellipticity in the \Hae\ photometry (bottom panel of Figure~\ref{Fig:H15_photometry}). When reporting the best-fit axis ratios below, we use the values computed from a luminosity-weighted average of the spatially varying axis ratios across the galaxy.

We perform model selection by searching through a large suite of galaxy models in the 6D parameter space described by the mass parameters $\mbh, \mlgen$, \mdm, and the three shape parameters.  For each galaxy model, a pair of libraries of stellar orbits sampling the phase space are integrated and their contributions to the stellar LOSVDs and light distribution are calculated.  A linear combination of the orbits in the libraries with non-negative weights is found that best represents the observed stellar kinematic moments and light distribution. A goodness-of-fit measure based on the log likelihood is then calculated for this model. To perform model selection and parameter inference, we iteratively approximate the log-likelihood landscape using Gaussian process regression and then populate the high-likelihood region with additional models. The dynamic nested sampler \texttt{dynesty} \citet{Speagleetal2020DYNESTY} is used to produce Bayesian posteriors for each log-likelihood surface and a uniform prior on each parameter. 

Further details of the procedures used here for triaxial orbit modeling and parameter inference can be found in our studies of other massive galaxies (e.g., \citealt{quennevilleetal2022, pilawaetal2022, liepoldetal2023}). Similar to those studies, we use the first eight Gauss-Hermite moments to parameterize the stellar LOSVDs but also constrain the next four terms in the series, $h_9$ through $h_{12}$, to be zero so as to reduce possible non-physical behaviors in the LOSVDs (see Section~7.1 of \citealt{liepoldetal2020}).

\subsection{Best-fit Galaxy Parameters}\label{sec:H15_results}

\begin{figure}
\centering
\includegraphics[width=1.0\linewidth]{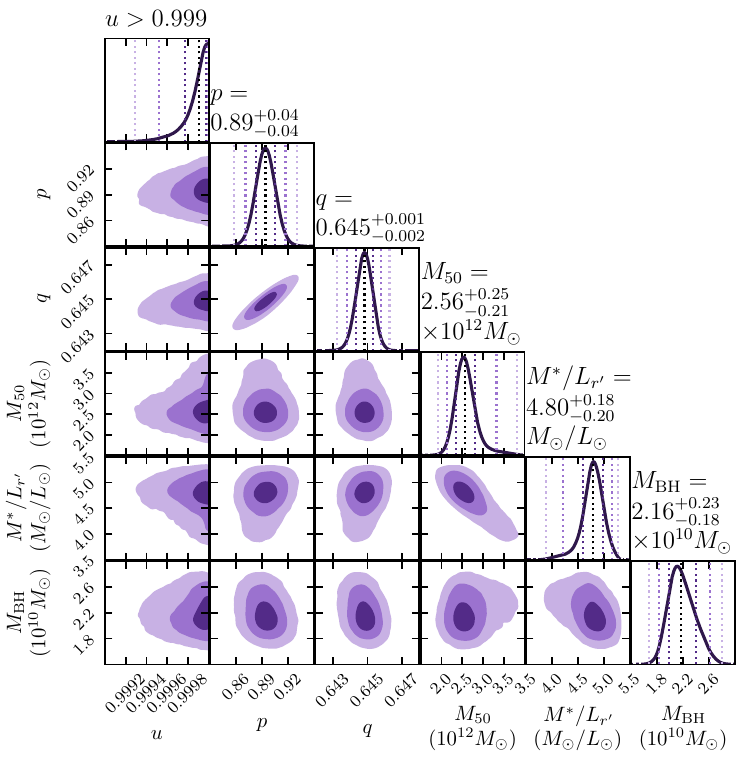}
\caption{
Posterior distributions of the six galaxy parameters in our triaxial Schwarzschild models of \Hae: black hole mass $\mbh$, stellar mass-to-light ratio $\mlgen_{r'}$, dark matter mass enclosed within 50 kpc \mdm, and luminosity-averaged axis ratios $p$, $q$, and $u$. 
The purple shaded regions mark the 68\%, 95\%, and 99.7\% confidence levels; the corresponding confidence levels in each 1D distribution are indicated by vertical dotted lines.
The listed uncertainties are 68\% intervals for the mass parameters and 99.7\% intervals for the axis ratios.
}
\label{Fig:H15_posterior}
\end{figure}

To determine the mass and shape parameters of \Hae, we begin with a preliminary round of parameter search, during which we compute stellar orbits for ${\sim}$15,000 TriOS galaxy models to coarsely sample a wide range in each of the six parameters as well as the second halo parameter $r_s$. This step enables us to broadly map out the goodness-of-fit landscape and to ensure sufficient coverage in the parameter space. After this initial survey, we compute another ${\sim}$23,000 galaxy models to construct the 6D landscape in higher fidelity. The resulting corner plot of the posterior distributions is displayed in Figure~\ref{Fig:H15_posterior} and the model parameters are summarized in Table~\ref{table:H15_best_fit_parameters}. A comparison between the $\chi^2$ of individual models and our inferred posterior distribution is shown in Appendix~\ref{appendix:posterior_comparison}.

The best-fit model has $\chi^2=1410$ and reproduces the observed stellar kinematics well, as illustrated in Figure~\ref{fig:H15_kinematic_profile}. The 6-parameter model has 2504 kinematic constraints, so a simplistic estimate of the reduced $\chi^2$ would yield $1410 / (2504-6) = 0.57$. In reality, however, the number of degrees of freedom (DOF) is often difficult or impossible to compute, in particular for nonlinear models such as here \citep[e.g.,][]{andraeetal2010}. This issue was examined in some detail in our analysis of NGC~2693 \citep{pilawaetal2022, pilawaetal2024}, where we estimated the number of ``generalized" DOF \citep{Ye1998, lipkathomas2021} for the best-fitting model to be $N_\mathrm{DOF} \sim 200$. Give that a total of 654 kinematic constraints were available for that galaxy, we would expect the inferred reduced $\chi^2$ to be raised by $(654-6)/(654-200)=1.43$ compared to the naive case of $N_\mathrm{DOF}=6$. Similarly for \Hae, we expect the reduced $\chi^2$ to also be larger when a similar estimate of the effective $N_\mathrm{DOF}$ is used.

\begin{table}
\begin{tabular}{l|l}
\Hae\ Galaxy Parameter (units)                                   & Inferred value \\ \hline
Black hole mass \mbh\ ($M_\odot$)                                & $(2.16_{-0.18}^{+0.23})\times 10^{10}$ \\ 
$\mlgen_{r'}$ ($M_\odot / L_{\odot}$)                            & $4.80_{-0.20}^{+0.18}$\\
$\mlgen_{V}$ ($M_\odot / L_{\odot}$)                             & $7.61_{-0.32}^{+0.29}$\\
Total stellar mass $M_*$ ($M_\odot$)                             & $(2.89_{-0.12}^{+0.11})\times 10^{12}$\\
Dark matter ($<50$ kpc) \mdm\ ($M_\odot$)                        & $\left( 2.56_{-0.21}^{+0.25} \right) \times 10^{12}$\\
Middle-to-long axis ratio $p$                                    & $0.89\pm0.04$ \\ 
Short-to-long axis ratio $q$                                     & $0.645_{-0.002}^{+0.001}$ \\
\multirow{1.8}{100pt}{Apparent-to-intrinsic long axis ratio $u$} & $>0.999$ \\
[-1ex]
& \\
Triaxiality $T$                                                  & $0.35\pm 0.03$ \\
\end{tabular}
\caption{
Triaxial orbit modeling results for the mass and shape parameters of \Hae.
The listed uncertainties are 68\% intervals for all quantities except the axis ratios, which are 99.7\% intervals.
}
\label{table:H15_best_fit_parameters}
\end{table}

The SMBH mass of \Hae\ is well constrained by our data: $\mbh = (2.16^{+0.23}_{-0.18})\times 10^{10} \msun$.  This SMBH is comparable in mass to that of NGC~4889, the BCG of the Coma cluster \citep{mcconnelletal2011a, mcconnelletal2012}. It is is also the most massive black hole in the local universe that has been modeled with a triaxial stellar component.  The $\mbh$ reported here, however, is nearly a factor of two lower than $\mbh = (4.0\pm 0.80)\times 10^{10} \msun$ obtained from axisymmetric orbit modeling of MUSE data in \citet{mehrganetal2019}. We have performed numerous tests and comparisons in an attempt to understand the difference and will discuss the results in Section~\ref{sec:mass_distribution_comparison}.

The stellar mass-to-light ratio of \Hae\ is also well constrained:  $\mlgen_{r'} = 4.80^{+0.18}_{-0.20} \, M_\odot / L_\odot$, corresponding to $\mlgen_V = 7.61^{+0.29}_{-0.32} \, M_\odot / L_\odot$ (for a color conversion of $V-r'=0.5$). This is consistent with the mean value of $\mlgen_{r} = 5.0\pm 1.2\  M_\odot / L_\odot$ (within the effective radius) inferred from stellar population synthesis modeling of 41 local early-type galaxies in the MASSIVE survey \citep{guetal2022}. Using the total luminosity determined in Section~\ref{sec:photometry}, we find the total stellar mass of \Hae\ to be $M_*=(2.89^{+0.11}_{-0.12}) \times 10^{12}\ \msun$.

With these parameters, the gravitational sphere of influence of the SMBH has a radius of $r_{\mathrm{SOI}} = 3.22\pm0.15\,\textrm{kpc} = 2.99''\pm0.15''$ when it is defined as $M_*(<r_\mathrm{SOI}) = \mbh$, and $r_{\mathrm {SOI}} = 4.31\pm0.22\,\textrm{kpc} = 4.00''\pm 0.20''$ when the alternative definition of $M_*(<r_\mathrm{SOI}) = 2 \mbh$ is used.

The enclosed dark matter mass (within 50 kpc) is reasonably constrained by our data: $M_{50}=(2.56^{+0.25}_{-0.21})\times 10^{12}\ \msun$,  corresponding to a central dark matter density of $\rho_0= (9.4^{+0.9}_{-0.8}) \times 10^6\ M_\odot\, \mathrm{kpc}^{-3}$.  As noted in Section~\ref{sec:triaxial_orbit_models}, we have set the halo scale radius to $r_s=150$ kpc. This choice is motivated by the findings from our initial coarse 7D parameter search that includes $r_s$: the range of $r_s \lesssim 50$ kpc is strongly disfavored (with $\chi^2$ at least 50 above the minimum).  In a suite of subsequent test models in which $r_s$ is fixed to 25 kpc, we find that the elevated $\chi^2$ are mainly due to the models' under-predictions of the kinematic moments $\sigma, h_4$ and $h_6$ beyond about 20 kpc, leading to ${\sim}30$\% lower dark matter halo mass than the best-fit model listed in Table~\ref{table:H15_best_fit_parameters}. Dark matter halos with small $r_s$ therefore appear to be too centrally concentrated to account for the observed kinematics.  On the other hand, when $r_s$ is allowed to be larger than ${\sim}50$ kpc (the outer reach of our data), the density profile of the halo within 50 kpc is essentially flat and does not depend strongly on the choice of $r_s$. For these halos, the kinematic data are insensitive to $r_s$ and mainly constrain the central dark matter density $\rho_0$, or equivalently, the enclosed dark matter $M_{50}$, as illustrated in Figure~\ref{Fig:H15_posterior}.

The 3D intrinsic shape of \Hae\ is constrained to be strongly triaxial and quite flattened, with a (luminosity-weighted) short-to-long intrinsic axis ratio of $q=0.65$. This low value of $q$ is commensurate with the flattened 2D isophotes of \Hae\ (Figure~\ref{Fig:H15_photometry}), where the apparent minor-to-major axis ratio on the sky, $q' = b'/a'=1-\epsilon$, is 0.75 at 10 kpc and 0.6 at 100 kpc. The tight correlation between $q$ and $p$ in Figure~\ref{Fig:H15_posterior} indicates the triaxiality parameter, $T = (1 - p^2)/(1 - q^2)$, is well constrained. Indeed, we find $T =0.35 \pm 0.03$ (i.e., an oblate ellipsoid), strongly ruling out both oblate axisymmetric ($T=0$) and prolate axisymmetric ($T=1$) shapes. This strong rejection of axisymmetry is consistent with the observed kinematic misalignment discussed in Section~\ref{sec:kinematics} as axisymmetric models cannot produce kinematic misalignment due to symmetry.

We recall that the axis ratios obey the inequality $q \le uq' \le p \le u \le 1$ \citep{quennevilleetal2022}. The preferred values of $q$ and $u$ are therefore both near their maximal allowed values ($u = 1$ and $q = u q'$). This limit corresponds to orientating \Hae's short and long axes on the sky plane and viewing it along the middle axis. This orientation corresponds to a nearly edge-on orientation for both short- and long-axis loop orbits. Overall, we find the inferred mass parameters not to change significantly as the viewing angles are moved away from the best-fit values. In a test in which the angles are fixed to ${>}15^\circ$ (${>}5\sigma$) away from the preferred range but the other parameters are allowed to vary, we find the best-fit \mbh\ to decrease by only 10\%.

The stellar orbits in the best-fit triaxial model are mildly anisotropic.  Using the standard anisotropy parameter, $\beta = 1 - \sigma_t^2 / \sigma_r^2$, where $\sigma_r$ and $\sigma_t$ are the intrinsic stellar velocity dispersion in the radial and tangential directions, respectively, we find the orbits to be somewhat more tangential ($\beta \approx -0.25$ and $\sigma_t/\sigma_r \approx 1.12$) in the central region and radial ($\beta \approx 0.25$ and $\sigma_t/\sigma_r \approx 0.87$) at larger radii. This transition from tangentially favored to radially favored orbital structure is similar to the trends seen in other massive cored galaxies with ultramassive SMBHs such as NGC~4889 and NGC~1600 \citep{mcconnelletal2011a, thomasetal2016}, but the radial gradient of $\beta$ in \Hae\ is milder.

The parameter uncertainties shown in Figure~\ref{Fig:H15_posterior} are obtained from our full analysis of the log-likelihood surface in the 6D parameter space. As a rough consistency check on the size of the uncertainties, we split our kinematic data into halves, once across the photometric major axis and once across the minor axis. For each of these four halves, we rerun the 6D parameter search using the reduced dataset as constraints. We find the four halves to select $\mbh = (2.15,\,2.39,\,2.59,\,1.82)\times10^{10}\msun$, giving $\mbh = (2.23\pm 0.29)\times 10^{10}\msun$, which is consistent with the value from our posterior distribution. Similarly, this test finds $\mlgen_{r'} = 4.64\pm0.11 \, M_\odot / L_\odot$ and $M_{50}=(2.6\pm0.2)\times 10^{12}\ \msun$, again fully consistent with the posterior-based results.

\subsection{\texorpdfstring{Scaling Relations between \mbh\ and Galaxy Properties}{Scaling Relations between the Supermassive Black Hole Mass and Galaxy Properties}} \label{sec:scaling}

\begin{figure}
\centering
\includegraphics[width=0.9\linewidth]{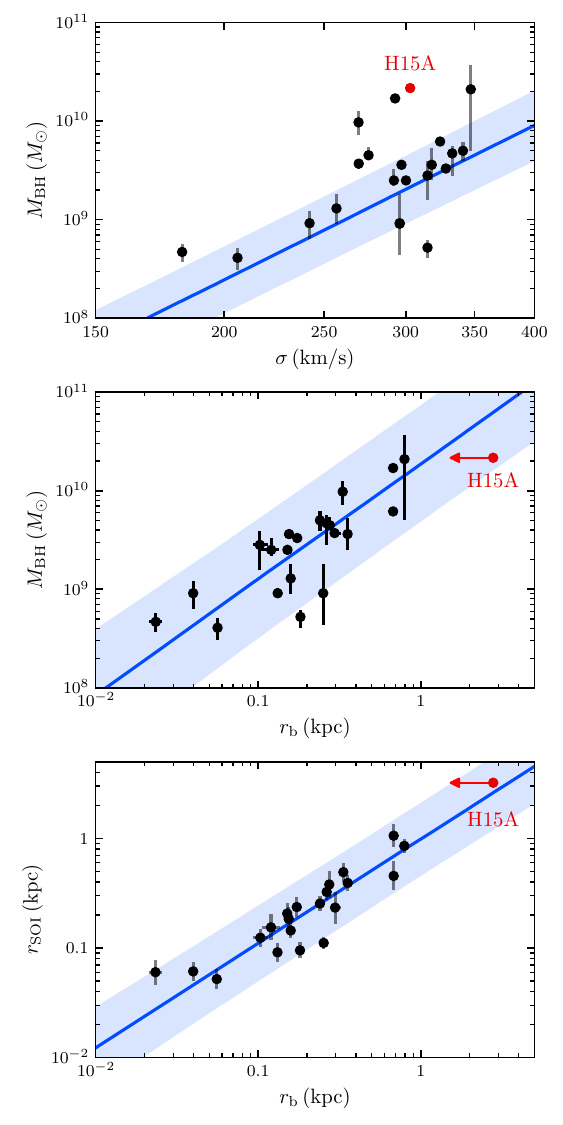}
\caption{
Scaling relations between \mbh\ and galaxy properties for local massive galaxies with dynamically determined \mbh\ and photometrically measured stellar core sizes \citep{thomasetal2016}.
(Top) \Hae\ (red circle) is a $3\sigma$ outlier above the mean $\mbh\text{--}\sigma$ relation for early-type galaxies.
(Middle) \Hae\ is ${\sim} 1\sigma$ below the mean relation between \mbh\ and host galaxy core radius $r_b$.
(Bottom) \Hae\ is near the mean relation between $r_{\rm SOI}$ and $r_b$.
Since some prior studies are able to fit \Hae's light profile without a core parameter, we indicate this uncertainty with a leftward red arrow.
} 
\label{Fig:scaling_relations}
\end{figure}

With our measurement of $\mbh=(2.16^{+0.23}_{-0.18})\times 10^{10}\msun$, \Hae\ -- along with NGC~4889 -- host the most massive SMBHs known in the local universe; furthermore, the uncertainty in \mbh\ is much smaller for \Hae\ than NGC~4889 \citep{mcconnelletal2011a, mcconnelletal2012}. Similar to NGC~4889, the \Hae\ SMBH is a significant outlier in the $\mbh\text{--}\sigma$ relation. Using the luminosity-weighted stellar velocity dispersion within $R_e=35\farcs4$ computed over our data, $\sigma = 303$ \kms, we find \mbh\ of \Hae\ to be a factor of 10.1 above the mean relation for early-type galaxies from \citet{mcconnellma2013} and a $3\sigma$ outlier (top panel of Figure~\ref{Fig:scaling_relations}).  At \Hae's total stellar mass of $M_*=(2.89^{+0.11}_{-0.12})\times 10^{12}\msun$, its \mbh\ is a factor of 2.5 above the mean $\mbh\text{--}M_{\rm bulge}$ relation of \citet{mcconnellma2013}.

To examine the correlation between \Hae's stellar core properties and \mbh, we recall that the reported core size of \Hae\ has ranged from zero to ${\sim}5$ kpc depending on the functional form assumed to approximate the surface brightness profile (Section~\ref{sec:intro}).  We perform a 2D fit over the GMOS image with a core-\sersic\ plus \sersic\ model using \imfit\ \citep{erwin2015} and find $n = 0.79$ and $r_b = 2.59'' = 2.79$ kpc for the core-\sersic\ component. These values are similar to those found in \citet{mehrganetal2019}.  Figure~\ref{Fig:scaling_relations} shows the location of \Hae\ on the $r_b\text{--}\mbh$ relation (middle panel) and $r_b\text{--}r_{\rm SOI}$ relation (bottom panel) for a sample of local massive early-type galaxies with both $r_b$ and \mbh\ measurements presented in \citet{thomasetal2016}.  Similar to these galaxies, the core size of \Hae\ is a more robust estimator than $\sigma$ is as an estimator for its \mbh.

\section{Comparison to Prior Work} \label{sec:discussion}

\begin{figure}
\centering
\includegraphics[width=1.\linewidth]{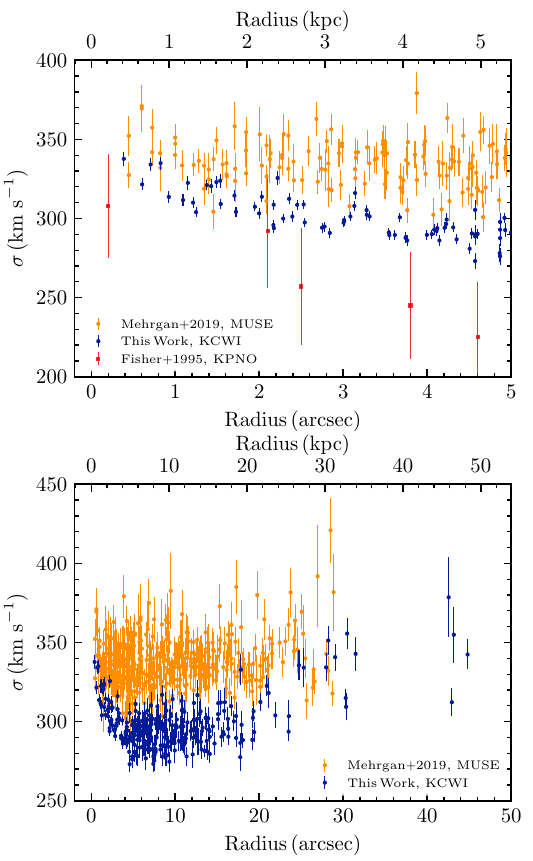}
\caption{
Radial profile of the stellar velocity dispersion of \Hae\ from KCWI (blue; this work), MUSE (orange; \citealt{mehrganetal2019}), and a long slit observation 
(red; PA $=-23^\circ$, \citealt{Fisheretal1995}).
The top panel shows the innermost $5''$; the bottom panel shows all available data out to a radius of $50''$.
}
\label{fig:H15_sigma_profile}
\end{figure}

\subsection{Stellar Kinematics: KCWI vs. MUSE} \label{sec:stellar_kinematic_comparison}

\citet{mehrganetal2019} used MUSE on the Very Large Telescope to observe the central $60''\times 60''$ region of H15. After spatial masking, their coverage is a roughly octagonal region with the outermost kinematic bins lying $28''$ from the center.  As illustrated in Figure~\ref{fig:H15_sigma_profile}, the radial profile of their $\sigma$ has a very different shape from ours: the MUSE $\sigma$ values in the central region (top panel) do not vary appreciably with radius and stay around a mean of 335 \kms\ while the KCWI $\sigma$ drops from 340 \kms\ in the innermost bin to 280-300 \kms\ at $R=5''$. Between $R=5''$ and $30''$, the mean value is $\sigma=295$ \kms\ for KCWI and $\sigma=339$ \kms\ for MUSE. The range $R=30''$ to $50''$ is covered only by KCWI data, where $\sigma$ rises back to $\approx 350$ \kms. For $h_4$, the mean value computed over $R=5''-30''$ is $h_4=0.037$ for KCWI and $h_4=0.086$ for MUSE.

To determine possible causes for the discrepancies between the KCWI and MUSE kinematics, we have identified three major differences between the two data analysis procedures and performed a series of tests on the KCWI data in an attempt to isolate the factors. Each difference is discussed below: (i) stellar templates, (ii) spectral fitting parameters, and (iii) spectral coverage. 

\subsubsection{Stellar Templates}

For the stellar templates used in the spectral fitting procedure, \citet{mehrganetal2019} adopt 16 stars from the MILES library \citep{sanchez-blazquezetal2006} at 2.5 \AA\ FWHM spectral resolution \citep{falcon-barrosoetal2011}. We instead use the higher-resolution stellar templates from the Indo-US library \citep{valdesetal2004}, which have a comparable spectral resolution (1.35 \AA\ FWHM) as our KCWI slicer data (Section~\ref{sec:reduction}). If the MILES templates were used, the small slicer data would have to be degraded to a worse resolution. But as a test, we repeat our spectral fitting process on the KCWI large slicer data using a similar subset of 16 templates from the MILES library as \citet{mehrganetal2019}. We find our stellar moments to be nearly identical to those obtained from the Indo-US templates. Stellar template choices therefore do not impact the KCWI kinematics for \Hae\ and do not remove the discrepancy.

\subsubsection{Spectral Fitting Parameters}
\citet{mehrganetal2019} truncate the Gauss-Hermite series representation of the stellar LOSVDs at $h_4$ instead of $h_8$. Repeating our KCWI analysis using only the lowest four moments leads to very small changes in $\sigma$ and $h_4$, with a mean difference of $\Delta\sigma=2.1$ \kms\ and $\Delta h_4=0.01$  This is not surprising because the amplitudes of the higher moments $h_5$ to $h_8$ in our fiducial analysis are all near 0 (Figure \ref{fig:H15_kinematic_profile}). We also test the robustness of our extracted KCWI stellar kinematics to the order of the multiplicative polynomial used to model the stellar continuum. There is negligible impact when the order is changed from $n=7$ to 5 or 9.

\subsubsection{Spectral Coverage}
A major difference between KCWI and MUSE is the spectral coverage: we use rest-frame $3885-5260$ \AA\ from KCWI, where the key absorption features are Ca H and K, the G-band, Fe lines, and \mgb\ (Figure~\ref{fig:spectra}); \citet{mehrganetal2019} use rest-frame $4750-6680$ \AA\ from MUSE, where the main feature is \mgb\ (other lines such as H$\beta$ and the 5270 \AA\ Fe are contaminated by gas emission or sky features). In an attempt to mimic the MUSE analysis, we use only the redder part of each KCWI spectrum that overlaps with the MUSE coverage (rest-frame $4750-5260$ \AA) and repeat our spectral fitting using only the first four Gauss-Hermite moments and the MILES templates.\footnote{This test is performed only for the KCWI large slicer data since the small slicer has a higher spectral resolution than MILES.} This test is repeated for three choices of the multiplicative polynomial order ($n=3$, $4$, and $5$). In the end, we find the mean $\sigma$ to change by no more than ${\sim}2$\% from the fiducial $\sigma$ in all cases. The $h_4$ moments show larger changes, with the mean $h_4$ increasing from the fiducial $h_4 = 0.031$ by $\Delta h_4=0.018$, $0.023$, and 0.018 for $n=3$, $4$, and $5$, but the increase is insufficient to account for the difference between KCWI and MUSE.

In summary, we find the KCWI kinematics to be very stable across the suite of tests designed to modify our analysis to mimic the MUSE study. While we obtain very similar $\sigma$ from fitting the full KCWI spectral range versus the \mgb\ region alone, we caution that prior studies based on other spectrographs have found it difficult to obtain robust $\sigma$ measurements using only the \mgb\ absorption lines. \citet{barthetal2002}, e.g., find up to 25\% larger $\sigma$ from fitting the  \mgb\ region than the more robust Ca triplet region. \citet{murphyetal2011} compare fits to five narrow windows centered around Ca H and K, G-band, H$\beta$, Fe, and the \mgb\ region, finding consistent $\sigma$ values from the first four regions but deviant $\sigma$ from the \mgb\ region. Since \mgb\ is the only dominant line in the MUSE spectra, further tests on MUSE data would be useful.

\subsubsection{Parametric vs. Non-parametric LOSVDs}

While \citet{mehrganetal2019} determined the LOSVDs from their spectra using their own non-parametric code, they also examined LOSVDs obtained from the popular parametric code pPXF (used in this work). The first four Gauss-Hermite moments of both sets of LOSVDs are compared in their Figure~18. The histograms in that figure show that there is no systematic difference in the moments obtained from the two methods. We have repeated the analysis using only their inner bins with $R < 5''$ and find very similar results. The methods used to extract the LOSVDs therefore do not appear to be the cause of the different results between our and their studies. 

Parametric vs. non-parametric method aside, we note that in neither study of \Hae\ is the LOSVD required to be positive definite at all velocities, in contrast to some prior studies that have explicitly imposed this criterion (e.g., \citealt{gebhardtetal2000a, pinkneyetal2003, falconbarrosomartig2021}). The Gauss-Hermite parametrization in general allows for LOSVDs that are negative at some velocities. We have examined the subset of our LOSVDs that have negative values at some velocity bins. For these, we find that even at the most negative location in the LOSVD, the mean ratio of the amplitude and the uncertainties at that velocity is only $-0.88$, indicating that the negativity is consistent with 0 within measurement uncertainties.

\subsection{Stellar Dynamical Models: Axisymmetry vs. Triaxiality} \label{sec:mass_distribution_comparison}

Stellar kinematics aside, another major difference between this study and \citet{mehrganetal2019} is the symmetry assumed for the stellar component in the orbit models: we allow for triaxiality while \citet{mehrganetal2019} assume axisymmetry.

Since \Hae\ exhibits prominent kinematic misalignment (Section~\ref{sec:kinematics} and Figure~\ref{fig:H15_vtheta_fit}) that cannot be produced by axisymmetric orbit models, relaxed assumptions about the stellar halo's intrinsic shape are essential for a proper representation of \Hae. Triaxial models, for example, are able to reproduce these types of kinematic features. Nonetheless, for comparison purposes, we perform a full round of orbit modeling with the axisymmetrized version of the TriOS code \citep{liepoldetal2020, quennevilleetal2021} and the same Keck data in order to quantify the effect of the assumed symmetry on \mbh\ and other parameters of \Hae.

\begin{figure}
\centering
\includegraphics[width=1.0\linewidth]{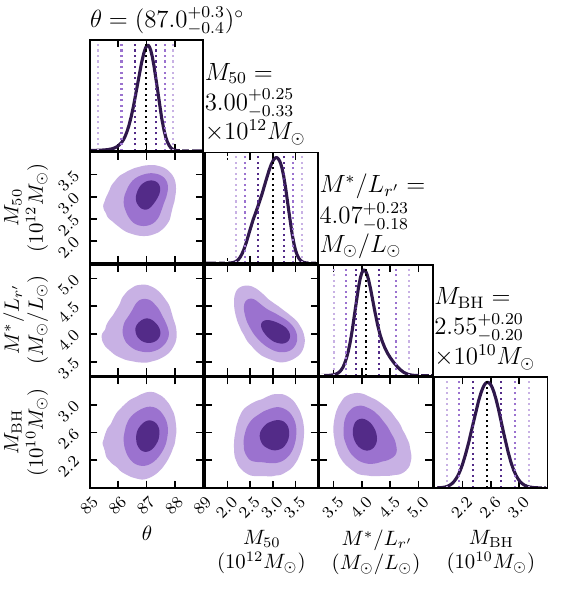}
\caption{
Posterior distributions of the four parameters in our axisymmetric orbit models of \Hae.
The mass parameters are the same as in Figure~\ref{Fig:H15_posterior};
$\theta$ is the inclination angle of an axisymmetric system.
The short-to-long axis ratio associated with this inclination is $q = 0.646\pm0.002$.
}
\label{Fig:axi_cornerplot}
\end{figure}

We compute ${\sim}$15,000 axisymmetric galaxy models to search a 4D parameter space of $\mbh, \mlgen_{r'}, \mdm$, and the inclination angle $i$. Figure~\ref{Fig:axi_cornerplot} shows the resulting posterior distributions.  The preferred mass parameters differ from the triaxial values by up to  $\sim 20$\%: the black hole mass increases by 18\% to $\mbh=  (2.55 \pm 0.20)\times 10^{10}\ \msun$; the stellar mass-to-light ratio decreases by 15\% to $\mlgen_{r'}=4.10\pm 0.21\ \msun/L_\odot$; the dark matter halo mass increases by 16\% to $M_{50}= (3.0\pm 0.3)\times 10^{12}\ \msun$.  The preferred inclination is nearly edge-on with $i = (87.0\pm 0.4)^\circ$. The best-fitting axisymmetric model is substantially less well fit than the best triaxial model, with a $\chi^2$ that is 330 higher.

While \mbh\ from our axisymmetric modeling is larger than our triaxial value, it is still significantly below the mass reported in \citet{mehrganetal2019}. Their $\mlgen$ is $4.75\pm 0.20$ when converted to the $r'$-band (using $i - r' \sim 0.06$ needed to match their and our photometry), which is 16\% higher than our value from axisymmetric modeling. For the dark matter halo, their enclosed mass within the MUSE FOV ($R\la 30$ kpc) is $M_{30} =1.4 \times 10^{12}\ \msun$, while our value is $M_{30}=8 \times 10^{11}\ \msun$. The inclination angle in their models is assumed to be $i=90^\circ$ (edge-on), which they justify using the flatness of \Hae's light distribution. Our preferred inclination $i=(87.0^{+0.3}_{-0.4})^\circ$ is close to this value but interestingly disfavors $90^\circ$ strongly. Since \mbh\ shows weak covariance with $i$ (Figure~\ref{Fig:axi_cornerplot}), the difference of $\sim 3^\circ$ is unlikely to be the cause of the factor of $\sim 2$ difference in \mbh.  

Most prior \mbh\ measurements based on stellar dynamical orbit modeling have assumed either a spherical or an axisymmetric stellar potential. In the handful of galaxies for which axisymmetric and triaxial modeling has been performed on the same data, the derived black hole masses are similar in some cases and differ in others. For instance, \mbh\ from axisymmetric modeling is smaller by 65\% for NGC~3379 \citep{vandenboschdezeeuw2010}, while for PGC~046832 \citep{denBroketal2021} and M59-UCD3 \citep{ahnetal2018}, axisymmetric modeling yields constraints on \mbh\ while triaxial modeling only provides an upper limit on \mbh. However, we caution that the triaxial results in these studies used the original Schwarzschild orbit code  \citep{vandenboschetal2008}, which has been shown to have 12 wrong signs when the loop orbits in one octant of the space are mirrored into the other 7 octants \citep{quennevilleetal2022}.  While correcting the mirroring errors still does not lead to a detection of SMBH in the case of PGC~046832 \citep{thateretal2022}, each galaxy should be re-examined separately. For two other galaxies where the TriOS code (with corrected orbital mapping) is used, axisymmetric modeling yields 40\% higher \mbh\ than triaxial modeling for NGC~2693 \citep{pilawaetal2022}, while the two methods give similar \mbh\ for NGC~1453 \citep{liepoldetal2020, quennevilleetal2022}. Our finding of a 18\% overestimate in \mbh\ from axisymmetric orbit modeling of \Hae\ is in line with these results.

\section{Summary}

Exploiting the superb sensitivity, stability, and efficiency of Keck KCWI over five observing runs between 2018 and 2021, we have obtained high-quality spatially-resolved spectra covering a ${\sim} 100''\times 100''$ FOV of \Hae. This remarkable galaxy has an exceptionally low central surface brightness (Figure \ref{Fig:H15_photometry}) and has been speculated and reported to harbor an unusually massive central SMBH. We co-add thousands of spectra from individual KCWI spaxels to form high-$S/N$ spectra for 313 contiguous spatial bins of H15 (Figure~\ref{fig:spectra}).  From each spectrum, we measure the full shape of the stellar LOSVD using eight Gauss-Hermite moments.  The pronounced misalignment between the photometric and kinematic axes of \Hae\ (Figure~\ref{fig:H15_vtheta_fit}) indicates the standard assumption of axisymmetry must be relaxed.  We instead perform full-scale triaxial Schwarzschild orbit modeling, using the KCWI stellar kinematics to constrain simultaneously the mass and shape parameters of \Hae.

Our results indicate that \Hae\ hosts one of the two most massive SMBHs known in the local universe and has an intrinsically triaxial stellar halo. The stellar mass-to-light ratio is consistent with expectations from stellar population synthesis models of massive ETGs, and the total stellar mass is among the largest known from dynamical methods. \Hae\ is a $3\sigma$ outlier in the $\mbh\text{--}\sigma$ relation, but it lies within the $1\sigma$ uncertainties in the $\mbh\text{--}M_{\rm bulge}$, $\mbh\text{--}r_b$, and $r_{\mathrm{SOI}}\text{--}r_\mathrm{b}$ relations.
At the highest masses, \citet{laueretal2007} show that galaxy velocity dispersions and luminosities predict inconsistent \mbh\ when the canonical $\mbh\text{--}\sigma$ and $\mbh\text{--}L$ relations are used.  Since galaxy $\sigma$ tends to saturate as galaxies merge (e.g., \citealt{boylankolchinetal2006}), they suggest galaxy light (or mass) as a better \mbh\ indicator and advocate direct dynamical measurements of \mbh\ in the most massive galaxies.  We have done such and have shown that \Hae\ is an example of local massive galaxies for which its velocity dispersion is a much poorer indicator of its dynamically measured \mbh\ than its stellar mass or core radius.

The low surface brightness at the centers of massive elliptical galaxies has been a main challenge in ongoing efforts searching for ultra-massive SMBHs of several billion solar masses and beyond. This work has demonstrated the capability of discovering these SMBHs afforded by sensitive IFUs on 10-meter-class ground-based telescopes such as Keck's KCWI.

\section*{Acknowledgments}

We thank John Blakeslee for processing the archival GMOS photometry used in the work. We are grateful to the referee for insightful comments.   E.R.L. and C.-P.M. acknowledge support of NSF AST-2206307, the Heising-Simons Foundation, and the Miller Institute for Basic Research in Science.  J.L.W. acknowledges support of NSF AST-2206219.  The spectroscopic data presented in this paper were obtained at the W. M. Keck Observatory, which is operated as a scientific partnership among the California Institute of Technology, the University of California and the National Aeronautics and Space Administration. The Observatory was made possible by the generous financial support of the W. M. Keck Foundation. 
This work used observations made with the NASA/ESA Hubble Space Telescope, obtained at the Space Telescope Science Institute, which is operated by the Association of Universities for Research in Astronomy, Inc., under NASA contract NAS5-26555.
This work used Expanse at the San Diego Supercomputing Center through allocation AST180041 from the Advanced Cyberinfrastructure Coordination Ecosystem: Services \& Support (ACCESS) program, which is supported by National Science Foundation grants \#2138259, \#2138286, \#2138307, \#2137603, and \#2138296.
\\[2ex] 

\vspace{5mm}
\facilities{Keck (KCWI)}

\software{
\texttt{TriOS} \citep{quennevilleetal2021,quennevilleetal2022},
orbit code \citep{vandenboschetal2008},
\texttt{dynesty} \citep{Speagle_2020},
\texttt{scikit-learn} \citep{scikit-learn},
\texttt{LHSMDU} \citep{deutschdeutsch2012,LHSMDU},
\texttt{pPXF} \citep{Cappellari2017},
\texttt{vorbin} \citep{cappellaricopin2003},
\texttt{MGE} \citep{Cappellari2002},
KCWI Data Reduction Pipeline \citep{morrisseyetal2018},
\imfit\ \citep{erwin2015},
Photutils \citep{photutils}
}

\appendix

\section{Kinematics of the best fit model of H15} \label{appendix:kinematic_profiles}

\begin{figure*}
\centering
\includegraphics[width=\linewidth]{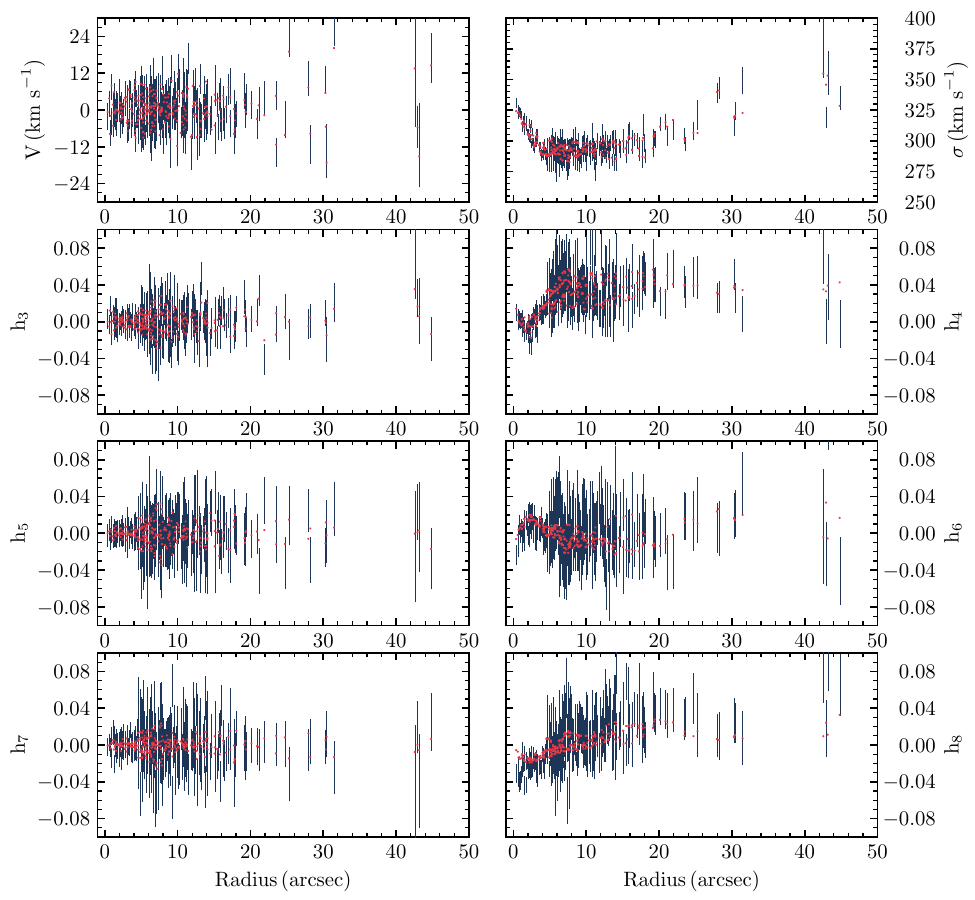}
\caption{
Radial profiles of the eight Gauss-Hermite moments of the stellar LOSVD for each of the 313 spatial bins of \Hae.  The black bars denote the values measured from KCWI stellar spectra, while red dots denote the values from the best-fitting triaxial \TriOS\ model.
The models produce point-symmetric stellar kinematics, so the data (black bars) shown here have been point-symmetrized for ease of comparison.
}
\label{fig:H15_kinematic_profile}
\end{figure*}

Radial profiles of the stellar kinematics used in the orbit modeling in this work and the best-fit model's reproduction of those kinematics are shown in Figure~\ref{fig:H15_kinematic_profile}.

\section{$\chi^2$ surface and marginalized posterior distribution}\label{appendix:posterior_comparison}

\begin{figure}
\centering
\includegraphics[width=\linewidth]{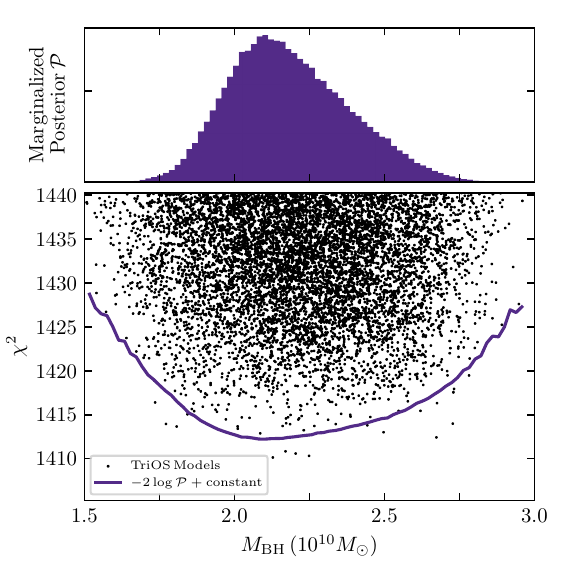}
\caption{
The marginalized posterior distribution from triaxial Schwarzschild models of \Hae\ (top), the log marginalized posterior (purple, lower), and the $\chi^2$ from individual \TriOS\ models (black dots) as a function of black hole mass. The marginalized posterior distribution is produced by modeling the $\chi^2$ surface over the six-dimensional parameter-space, then performing dynamic nested sampling that surrogate function.}
\label{fig:posterior_comparison}
\end{figure}

Figure~\ref{fig:posterior_comparison} shows our posterior distribution against the distribution of $\chi^2$ vs. the assumed black hole mass for the individual models in our analysis in Several prior works have used these sorts of distributions to determine their reported parameter values and uncertainties  \citep[e.g.,][]{gebhardtetal2011, walshetal2016}. While this type of analysis is reasonably accurate for low-dimensional searches with roughly Gaussian posteriors, it becomes increasingly inexact as the dimensionality rises and the posteriors deviate from Gaussian. In this high-dimensional space with non-Gaussian posteriors (see Figure~\ref{Fig:H15_posterior}), the marginalized posterior need not perfectly trace the lower contour of models in the $\chi^2$ vs \mbh\ space, though the two should still have an approximate correspondence. We find that the distributions of the individual $\chi^2$ values are broadly consistent with the center and width of our posterior for each parameter.

\section{Multi-Gaussian Expansion Fits} \label{appendix:mge}

We parameterize the surface brightness of \Hae\ using the \texttt{mgefit} routine \cite{Cappellari2017}. The image is represented by seven Gaussian components, each with its own width $\sigma'$, axis ratio $q'$, and central surface brightness $I$, but sharing a common center and a fixed PA of $-34^\circ$.  The fits reproduce the observations well, with residuals of less than 10\% for sectors within $100''$ of the center. The parameters for these components are listed in Table~\ref{H15_MGE_parameters}. The MGE is in the GMOS \emph{r\_G0326} band (similar to SDSS $r'$) and the surface brightness has been corrected for cosmological dimming and galactic extinction ($A_{r'} = 0.086$).

The smallest $\sigma'$ allowed in the MGE fit is $0.96''$. As previously discussed in \cite{liepoldetal2020}, we find that some configurations of the \texttt{mgefit} fitting routine produce centrally peaked deprojected light distributions. This occurs because the width of the innermost component of the fit is poorly constrained below the scale of the PSF. On the other hand, after deprojection the central luminosity density scales with $\nu_{0}\propto \Sigma_0/\sigma'$ (where $\Sigma_0$ is the central surface brightness). We find that often $\sigma' \ll \sigma'_{\textrm{PSF}}$ is selected by the fit, resulting in exceptionally large and unconstrained central densities. We find that placing a constraint on the fit that $\sigma' > 0.96''$ balances the quality of fit to the centralmost pixels and the shape of the deprojected light profile.

The selected axis ratios of the individual components roughly trace the ellipticity of the light profile described in Section~\ref{sec:photometry}, with $q' \sim 1 - \epsilon = 1.000$ for the innermost component and $q' = 0.516$ for the outermost component with $\sigma' = 140''$. In preliminary fits, the best-fit MGE model contained two components of disparate flattening at large radii: one with $\sigma' = 140''$ and $q' = 0.25$, and one with $\sigma' = 84.3''$ with $q' = 1.00$. In comparison, the isophotal shape based on our \texttt{ellipse} fit to the Gemini imaging data (see Section~\ref{sec:photometry}) is $q' = 1 - \epsilon \sim 0.62$ in this range of radii. While this set of parameters reproduced the observed data, the extreme flattening of that outermost component substantially restricts the deprojectability of the MGE. To resolve this issue, we instead compute the contribution to the MGE model from the outermost four components and then refit that contribution with only three components. The resulting fit yields similar residuals to the data as the original fit, and $q'$ varies smoothly with radius as seen in Table~\ref{H15_MGE_parameters}.

\begin{table}
\vspace{2ex}
\begin{tabular}{c|c|c}
\hline
$I_k \,[L_{\odot,r'} / {\mathrm{pc}}^{2}]$ & $ \ \ \sigma_k^{\prime} \ \ [\mathrm{arcsec}] \ \ $ & $ \ \ q_k^{\prime} \ \ $ \\
\hline
$59.447$ & $0.960$ & $1.000$ \\
$130.89$ & $3.053$ & $0.998$ \\
$116.98$ & $6.723$ & $0.725$ \\
$104.73$ & $11.44$ & $0.795$ \\
$28.947$ & $23.32$ & $0.669$ \\
$8.7178$ & $45.59$ & $0.627$ \\
$2.1461$ & $140.0$ & $0.516$ \\
\end{tabular}
\centering
\caption{
MGE parameters for the surface brightness of \Hae, after the modifications described in Section~\ref{appendix:mge}.
For each of the 7 two-dimensional Gaussian components, the first column lists the central surface brightness density, the middle column lists the dispersion of the Gaussian, and the last column lists the axis ratio, where primed variables denote projected quantities.
The surface brightness has been corrected for cosmological dimming and galactic extinction ($A_{r'} = 0.086$ mags). We use $m_{\odot,r'} = 4.65$ in the AB magnitude system.
}
\label{H15_MGE_parameters}
\end{table}

\bibliography{H15A}{}

\begin{thebibliography}{}
\expandafter\ifx\csname natexlab\endcsname\relax\def\natexlab#1{#1}\fi
\providecommand{\url}[1]{\href{#1}{#1}}
\providecommand{\dodoi}[1]{doi:~\href{http://doi.org/#1}{\nolinkurl{#1}}}
\providecommand{\doeprint}[1]{\href{http://ascl.net/#1}{\nolinkurl{http://ascl.net/#1}}}
\providecommand{\doarXiv}[1]{\href{https://arxiv.org/abs/#1}{\nolinkurl{https://arxiv.org/abs/#1}}}

\bibitem[{{Ahn} {et~al.}(2018){Ahn}, {Seth}, {Cappellari}, {Krajnovi{\'c}}, {Strader}, {Voggel}, {Walsh}, {Bahramian}, {Baumgardt}, {Brodie}, {Chilingarian}, {Chomiuk}, {den Brok}, {Frank}, {Hilker}, {McDermid}, {Mieske}, {Neumayer}, {Nguyen}, {Pechetti}, {Romanowsky}, \& {Spitler}}]{ahnetal2018}
{Ahn}, C.~P., {Seth}, A.~C., {Cappellari}, M., {et~al.} 2018, \apj, 858, 102, \dodoi{10.3847/1538-4357/aabc57}

\bibitem[{{Andrae} {et~al.}(2010){Andrae}, {Schulze-Hartung}, \& {Melchior}}]{andraeetal2010}
{Andrae}, R., {Schulze-Hartung}, T., \& {Melchior}, P. 2010, arXiv e-prints, arXiv:1012.3754.
\newblock \doarXiv{1012.3754}

\bibitem[{{Barth} {et~al.}(2002){Barth}, {Ho}, \& {Sargent}}]{barthetal2002}
{Barth}, A.~J., {Ho}, L.~C., \& {Sargent}, W.~L.~W. 2002, \aj, 124, 2607, \dodoi{10.1086/343840}

\bibitem[{{Begelman} {et~al.}(1980){Begelman}, {Blandford}, \& {Rees}}]{begelmanetal1980}
{Begelman}, M.~C., {Blandford}, R.~D., \& {Rees}, M.~J. 1980, \nat, 287, 307, \dodoi{10.1038/287307a0}

\bibitem[{{Bonfini} {et~al.}(2015){Bonfini}, {Dullo}, \& {Graham}}]{Bonfini2015}
{Bonfini}, P., {Dullo}, B.~T., \& {Graham}, A.~W. 2015, \apj, 807, 136, \dodoi{10.1088/0004-637X/807/2/136}

\bibitem[{{Boylan-Kolchin} {et~al.}(2006){Boylan-Kolchin}, {Ma}, \& {Quataert}}]{boylankolchinetal2006}
{Boylan-Kolchin}, M., {Ma}, C.-P., \& {Quataert}, E. 2006, \mnras, 369, 1081, \dodoi{10.1111/j.1365-2966.2006.10379.x}

\bibitem[{Bradley {et~al.}(2024)Bradley, Sipőcz, Robitaille, Tollerud, Vinícius, Deil, Barbary, Wilson, Busko, Donath, Günther, Cara, Lim, Meßlinger, Burnett, Conseil, Droettboom, Bostroem, Bray, Bratholm, Jamieson, Ginsburg, Barentsen, Craig, Pascual, Rathi, Perrin, Morris, \& Perren}]{photutils}
Bradley, L., Sipőcz, B., Robitaille, T., {et~al.} 2024, astropy/photutils: 1.13.0, 1.13.0,  Zenodo, \dodoi{10.5281/zenodo.12585239}

\bibitem[{Cappellari(2002)}]{Cappellari2002}
Cappellari, M. 2002, \mnras, 333, 400, \dodoi{10.1046/j.1365-8711.2002.05412.x}

\bibitem[{{Cappellari}(2017)}]{Cappellari2017}
{Cappellari}, M. 2017, \mnras, 466, 798, \dodoi{10.1093/mnras/stw3020}

\bibitem[{{Cappellari} \& {Copin}(2003)}]{cappellaricopin2003}
{Cappellari}, M., \& {Copin}, Y. 2003, \mnras, 342, 345, \dodoi{10.1046/j.1365-8711.2003.06541.x}

\bibitem[{{den Brok} {et~al.}(2021){den Brok}, {Krajnovi{\'c}}, {Emsellem}, {Brinchmann}, \& {Maseda}}]{denBroketal2021}
{den Brok}, M., {Krajnovi{\'c}}, D., {Emsellem}, E., {Brinchmann}, J., \& {Maseda}, M. 2021, \mnras, 508, 4786, \dodoi{10.1093/mnras/stab2852}

\bibitem[{Deutsch \& Deutsch(2012)}]{deutschdeutsch2012}
Deutsch, J.~L., \& Deutsch, C.~V. 2012, Journal of Statistical Planning and Inference, 142, 763, \dodoi{https://doi.org/10.1016/j.jspi.2011.09.016}

\bibitem[{{Edwards} {et~al.}(2016){Edwards}, {Alpert}, {Trierweiler}, {Abraham}, \& {Beizer}}]{edwardsetal2016}
{Edwards}, L.~O.~V., {Alpert}, H.~S., {Trierweiler}, I.~L., {Abraham}, T., \& {Beizer}, V.~G. 2016, \mnras, 461, 230, \dodoi{10.1093/mnras/stw1314}

\bibitem[{{Ene} {et~al.}(2018){Ene}, {Ma}, {Veale}, {Greene}, {Thomas}, {Blakeslee}, {Foster}, {Walsh}, {Ito}, \& {Goulding}}]{eneetal2018}
{Ene}, I., {Ma}, C.-P., {Veale}, M., {et~al.} 2018, \mnras, 479, 2810, \dodoi{10.1093/mnras/sty1649}

\bibitem[{{Erwin}(2015)}]{erwin2015}
{Erwin}, P. 2015, \apj, 799, 226, \dodoi{10.1088/0004-637X/799/2/226}

\bibitem[{{Faber} {et~al.}(1997)}]{faberetal1997}
{Faber}, S.~M., {et~al.} 1997, \aj, 114, 1771, \dodoi{10.1086/118606}

\bibitem[{{Falc{\'o}n-Barroso} \& {Martig}(2021)}]{falconbarrosomartig2021}
{Falc{\'o}n-Barroso}, J., \& {Martig}, M. 2021, \aap, 646, A31, \dodoi{10.1051/0004-6361/202039624}

\bibitem[{{Falc{\'o}n-Barroso} {et~al.}(2011){Falc{\'o}n-Barroso}, {S{\'a}nchez-Bl{\'a}zquez}, {Vazdekis}, {Ricciardelli}, {Cardiel}, {Cenarro}, {Gorgas}, \& {Peletier}}]{falcon-barrosoetal2011}
{Falc{\'o}n-Barroso}, J., {S{\'a}nchez-Bl{\'a}zquez}, P., {Vazdekis}, A., {et~al.} 2011, \aap, 532, A95, \dodoi{10.1051/0004-6361/201116842}

\bibitem[{{Ferrarese} \& {Merritt}(2000)}]{ferraresemerritt2000}
{Ferrarese}, L., \& {Merritt}, D. 2000, \apjl, 539, L9, \dodoi{10.1086/312838}

\bibitem[{Fisher {et~al.}(1995)Fisher, Illingworth, \& Franx}]{Fisheretal1995}
Fisher, D., Illingworth, G., \& Franx, M. 1995, \apj, 438, 539, \dodoi{10.1086/175100}

\bibitem[{{Gebhardt} {et~al.}(2011){Gebhardt}, {Adams}, {Richstone}, {Lauer}, {Faber}, {G{\"u}ltekin}, {Murphy}, \& {Tremaine}}]{gebhardtetal2011}
{Gebhardt}, K., {Adams}, J., {Richstone}, D., {et~al.} 2011, \apj, 729, 119, \dodoi{10.1088/0004-637X/729/2/119}

\bibitem[{{Gebhardt} {et~al.}(2000){Gebhardt}, {Richstone}, {Kormendy}, {Lauer}, {Ajhar}, {Bender}, {Dressler}, {Faber}, {Grillmair}, {Magorrian}, \& {Tremaine}}]{gebhardtetal2000a}
{Gebhardt}, K., {Richstone}, D., {Kormendy}, J., {et~al.} 2000, \aj, 119, 1157, \dodoi{10.1086/301240}

\bibitem[{{Gu} {et~al.}(2022){Gu}, {Greene}, {Newman}, {Kreisch}, {Quenneville}, {Ma}, \& {Blakeslee}}]{guetal2022}
{Gu}, M., {Greene}, J.~E., {Newman}, A.~B., {et~al.} 2022, \apj, 932, 103, \dodoi{10.3847/1538-4357/ac69ea}

\bibitem[{{Hoessel}(1980)}]{Hoessel1980}
{Hoessel}, J.~G. 1980, \apj, 241, 493, \dodoi{10.1086/158364}

\bibitem[{{Hoessel} {et~al.}(1980){Hoessel}, {Gunn}, \& {Thuan}}]{hoessel_paper_I_1980}
{Hoessel}, J.~G., {Gunn}, J.~E., \& {Thuan}, T.~X. 1980, \apj, 241, 486, \dodoi{10.1086/158363}

\bibitem[{{Holmberg}(1937)}]{Holmberg1937}
{Holmberg}, E. 1937, Annals of the Observatory of Lund, 6, 1

\bibitem[{{Jedrzejewski}(1987)}]{jedrzejewski1987}
{Jedrzejewski}, R.~I. 1987, \mnras, 226, 747, \dodoi{10.1093/mnras/226.4.747}

\bibitem[{{Kormendy} \& {Bender}(2009)}]{kormendybender2009}
{Kormendy}, J., \& {Bender}, R. 2009, \apjl, 691, L142, \dodoi{10.1088/0004-637X/691/2/L142}

\bibitem[{{Krajnovi{\'c}} {et~al.}(2006){Krajnovi{\'c}}, {Cappellari}, {de Zeeuw}, \& {Copin}}]{krajnovicetal2006}
{Krajnovi{\'c}}, D., {Cappellari}, M., {de Zeeuw}, P.~T., \& {Copin}, Y. 2006, \mnras, 366, 787, \dodoi{10.1111/j.1365-2966.2005.09902.x}

\bibitem[{{Lauer} {et~al.}(1995){Lauer}, {Ajhar}, {Byun}, {Dressler}, {Faber}, {Grillmair}, {Kormendy}, {Richstone}, \& {Tremaine}}]{laueretal1995}
{Lauer}, T.~R., {Ajhar}, E.~A., {Byun}, Y.~I., {et~al.} 1995, \aj, 110, 2622, \dodoi{10.1086/117719}

\bibitem[{{Lauer} {et~al.}(2005){Lauer}, {Faber}, {Gebhardt}, {Richstone}, {Tremaine}, {Ajhar}, {Aller}, {Bender}, {Dressler}, {Filippenko}, {Green}, {Grillmair}, {Ho}, {Kormendy}, {Magorrian}, {Pinkney}, \& {Siopis}}]{laueretal2005}
{Lauer}, T.~R., {Faber}, S.~M., {Gebhardt}, K., {et~al.} 2005, \aj, 129, 2138, \dodoi{10.1086/429565}

\bibitem[{{Lauer} {et~al.}(2007)}]{laueretal2007}
{Lauer}, T.~R., {et~al.} 2007, \apj, 662, 808, \dodoi{10.1086/518223}

\bibitem[{{Liepold} {et~al.}(2023){Liepold}, {Ma}, \& {Walsh}}]{liepoldetal2023}
{Liepold}, E.~R., {Ma}, C.-P., \& {Walsh}, J.~L. 2023, \apjl, 945, L35, \dodoi{10.3847/2041-8213/acbbcf}

\bibitem[{Liepold {et~al.}(2020)Liepold, Quenneville, Ma, Walsh, McConnell, Greene, \& Blakeslee}]{liepoldetal2020}
Liepold, E.~R., Quenneville, M.~E., Ma, C.-P., {et~al.} 2020, \apj, 891, 4, \dodoi{10.3847/1538-4357/ab6f71}

\bibitem[{{Lipka} \& {Thomas}(2021)}]{lipkathomas2021}
{Lipka}, M., \& {Thomas}, J. 2021, \mnras, \dodoi{10.1093/mnras/stab1092}

\bibitem[{{L{\'o}pez-Cruz} {et~al.}(2014){L{\'o}pez-Cruz}, {A{\~n}orve}, {Birkinshaw}, {Worrall}, {Ibarra-Medel}, {Barkhouse}, {Torres-Papaqui}, \& {Motta}}]{lopez-cruzetal2014}
{L{\'o}pez-Cruz}, O., {A{\~n}orve}, C., {Birkinshaw}, M., {et~al.} 2014, \apjl, 795, L31, \dodoi{10.1088/2041-8205/795/2/L31}

\bibitem[{{Ma} {et~al.}(2014){Ma}, {Greene}, {McConnell}, {Janish}, {Blakeslee}, {Thomas}, \& {Murphy}}]{maetal2014}
{Ma}, C.-P., {Greene}, J.~E., {McConnell}, N., {et~al.} 2014, \apj, 795, 158, \dodoi{10.1088/0004-637X/795/2/158}

\bibitem[{{Madrid} \& {Donzelli}(2016)}]{madriddonzelli2016}
{Madrid}, J.~P., \& {Donzelli}, C.~J. 2016, \apj, 819, 50, \dodoi{10.3847/0004-637X/819/1/50}

\bibitem[{{McConnell} \& {Ma}(2013)}]{mcconnellma2013}
{McConnell}, N.~J., \& {Ma}, C.-P. 2013, \apj, 764, 184, \dodoi{10.1088/0004-637X/764/2/184}

\bibitem[{{McConnell} {et~al.}(2011){McConnell}, {Ma}, {Gebhardt}, {Wright}, {Murphy}, {Lauer}, {Graham}, \& {Richstone}}]{mcconnelletal2011a}
{McConnell}, N.~J., {Ma}, C.-P., {Gebhardt}, K., {et~al.} 2011, \nat, 480, 215, \dodoi{10.1038/nature10636}

\bibitem[{McConnell {et~al.}(2012)McConnell, Ma, Murphy, Gebhardt, Lauer, Graham, Wright, \& Richstone}]{mcconnelletal2012}
McConnell, N.~J., Ma, C.~P., Murphy, J.~D., {et~al.} 2012, \apj, 756, \dodoi{10.1088/0004-637X/756/2/179}

\bibitem[{{Mehrgan} {et~al.}(2019){Mehrgan}, {Thomas}, {Saglia}, {Mazzalay}, {Erwin}, {Bender}, {Kluge}, \& {Fabricius}}]{mehrganetal2019}
{Mehrgan}, K., {Thomas}, J., {Saglia}, R., {et~al.} 2019, \apj, 887, 195, \dodoi{10.3847/1538-4357/ab5856}

\bibitem[{{Morrissey} {et~al.}(2018){Morrissey}, {Matuszewski}, {Martin}, {Neill}, {Epps}, {Fucik}, {Weber}, {Darvish}, {Adkins}, {Allen}, {Bartos}, {Belicki}, {Cabak}, {Callahan}, {Cowley}, {Crabill}, {Deich}, {Delecroix}, {Doppman}, {Hilyard}, {James}, {Kaye}, {Kokorowski}, {Kwok}, {Lanclos}, {Milner}, {Moore}, {O'Sullivan}, {Parihar}, {Park}, {Phillips}, {Rizzi}, {Rockosi}, {Rodriguez}, {Salaun}, {Seaman}, {Sheikh}, {Weiss}, \& {Zarzaca}}]{morrisseyetal2018}
{Morrissey}, P., {Matuszewski}, M., {Martin}, D.~C., {et~al.} 2018, \apj, 864, 93, \dodoi{10.3847/1538-4357/aad597}

\bibitem[{Moza(2020)}]{LHSMDU}
Moza, S. 2020, {sahilm89/lhsmdu: Latin Hypercube Sampling with Multi-Dimensional Uniformity (LHSMDU): Speed Boost minor compatibility fixes}, 1.1.1,  Zenodo, \dodoi{10.5281/zenodo.3929531}

\bibitem[{{Murphy} {et~al.}(2011){Murphy}, {Gebhardt}, \& {Adams}}]{murphyetal2011}
{Murphy}, J.~D., {Gebhardt}, K., \& {Adams}, J.~J. 2011, \apj, 729, 129, \dodoi{10.1088/0004-637X/729/2/129}

\bibitem[{{Navarro} {et~al.}(1996){Navarro}, {Frenk}, \& {White}}]{Navarroetal1996}
{Navarro}, J.~F., {Frenk}, C.~S., \& {White}, S. D.~M. 1996, \apj, 462, 563, \dodoi{10.1086/177173}

\bibitem[{Pedregosa {et~al.}(2011)Pedregosa, Varoquaux, Gramfort, Michel, Thirion, Grisel, Blondel, Prettenhofer, Weiss, Dubourg, Vanderplas, Passos, Cournapeau, Brucher, Perrot, \& Duchesnay}]{scikit-learn}
Pedregosa, F., Varoquaux, G., Gramfort, A., {et~al.} 2011, Journal of Machine Learning Research, 12, 2825

\bibitem[{{Pilawa} {et~al.}(2024){Pilawa}, {Liepold}, \& {Ma}}]{pilawaetal2024}
{Pilawa}, J., {Liepold}, E.~R., \& {Ma}, C.-P. 2024, \apj, 966, 205, \dodoi{10.3847/1538-4357/ad3935}

\bibitem[{{Pilawa} {et~al.}(2022){Pilawa}, {Liepold}, {Delgado Andrade}, {Walsh}, {Ma}, {Quenneville}, {Greene}, \& {Blakeslee}}]{pilawaetal2022}
{Pilawa}, J.~D., {Liepold}, E.~R., {Delgado Andrade}, S.~C., {et~al.} 2022, \apj, 928, 178, \dodoi{10.3847/1538-4357/ac58fd}

\bibitem[{Pinkney {et~al.}(2003)Pinkney, Gebhardt, Bender, Bower, Dressler, Faber, Filippenko, Green, Ho, Kormendy, Lauer, Magorrian, Richstone, \& Tremaine}]{pinkneyetal2003}
Pinkney, J., Gebhardt, K., Bender, R., {et~al.} 2003, \apj, 596, 903, \dodoi{10.1086/378118}

\bibitem[{{Quenneville} {et~al.}(2021){Quenneville}, {Liepold}, \& {Ma}}]{quennevilleetal2021}
{Quenneville}, M.~E., {Liepold}, E.~R., \& {Ma}, C.-P. 2021, \apjs, 254, 25, \dodoi{10.3847/1538-4365/abe6a0}

\bibitem[{{Quenneville} {et~al.}(2022){Quenneville}, {Liepold}, \& {Ma}}]{quennevilleetal2022}
---. 2022, \apj, 926, 30, \dodoi{10.3847/1538-4357/ac3e68}

\bibitem[{{Rusli} {et~al.}(2013){Rusli}, {Thomas}, {Saglia}, {Fabricius}, {Erwin}, {Bender}, {Nowak}, {Lee}, {Riffeser}, \& {Sharp}}]{ruslietal2013}
{Rusli}, S.~P., {Thomas}, J., {Saglia}, R.~P., {et~al.} 2013, \aj, 146, 45, \dodoi{10.1088/0004-6256/146/3/45}

\bibitem[{{S{\'a}nchez-Bl{\'a}zquez} {et~al.}(2006){S{\'a}nchez-Bl{\'a}zquez}, {Peletier}, {Jim{\'e}nez-Vicente}, {Cardiel}, {Cenarro}, {Falc{\'o}n-Barroso}, {Gorgas}, {Selam}, \& {Vazdekis}}]{sanchez-blazquezetal2006}
{S{\'a}nchez-Bl{\'a}zquez}, P., {Peletier}, R.~F., {Jim{\'e}nez-Vicente}, J., {et~al.} 2006, \mnras, 371, 703, \dodoi{10.1111/j.1365-2966.2006.10699.x}

\bibitem[{{Speagle}(2020{\natexlab{a}})}]{Speagleetal2020DYNESTY}
{Speagle}, J.~S. 2020{\natexlab{a}}, \mnras, 493, 3132, \dodoi{10.1093/mnras/staa278}

\bibitem[{{Speagle}(2020{\natexlab{b}})}]{Speagle_2020}
---. 2020{\natexlab{b}}, \mnras, 493, 3132, \dodoi{10.1093/mnras/staa278}

\bibitem[{{Thater} {et~al.}(2022){Thater}, {Jethwa}, {Tahmasebzadeh}, {Zhu}, {den Brok}, {Santucci}, {Ding}, {Poci}, {Lilley}, {Tim de Zeeuw}, {Zocchi}, {Maindl}, {Rigamonti}, {Yang}, {Fahrion}, \& {van de Ven}}]{thateretal2022}
{Thater}, S., {Jethwa}, P., {Tahmasebzadeh}, B., {et~al.} 2022, \aap, 667, A51, \dodoi{10.1051/0004-6361/202243926}

\bibitem[{{Thomas} {et~al.}(2016){Thomas}, {Ma}, {McConnell}, {Greene}, {Blakeslee}, \& {Janish}}]{thomasetal2016}
{Thomas}, J., {Ma}, C.-P., {McConnell}, N.~J., {et~al.} 2016, \nat, 532, 340, \dodoi{10.1038/nature17197}

\bibitem[{{Thomas} {et~al.}(2014){Thomas}, {Saglia}, {Bender}, {Erwin}, \& {Fabricius}}]{thomasetal2014}
{Thomas}, J., {Saglia}, R.~P., {Bender}, R., {Erwin}, P., \& {Fabricius}, M. 2014, \apj, 782, 39, \dodoi{10.1088/0004-637X/782/1/39}

\bibitem[{{Valdes} {et~al.}(2004){Valdes}, {Gupta}, {Rose}, {Singh}, \& {Bell}}]{valdesetal2004}
{Valdes}, F., {Gupta}, R., {Rose}, J.~A., {Singh}, H.~P., \& {Bell}, D.~J. 2004, \apjs, 152, 251, \dodoi{10.1086/386343}

\bibitem[{{van den Bosch} \& {de Zeeuw}(2010)}]{vandenboschdezeeuw2010}
{van den Bosch}, R.~C.~E., \& {de Zeeuw}, P.~T. 2010, \mnras, 401, 1770, \dodoi{10.1111/j.1365-2966.2009.15832.x}

\bibitem[{{van den Bosch} {et~al.}(2008){van den Bosch}, {van de Ven}, {Verolme}, {Cappellari}, \& {de Zeeuw}}]{vandenboschetal2008}
{van den Bosch}, R.~C.~E., {van de Ven}, G., {Verolme}, E.~K., {Cappellari}, M., \& {de Zeeuw}, P.~T. 2008, \mnras, 385, 647, \dodoi{10.1111/j.1365-2966.2008.12874.x}

\bibitem[{{Vazdekis} {et~al.}(2016){Vazdekis}, {Koleva}, {Ricciardelli}, {R{\"o}ck}, \& {Falc{\'o}n-Barroso}}]{vazdekisetal2016}
{Vazdekis}, A., {Koleva}, M., {Ricciardelli}, E., {R{\"o}ck}, B., \& {Falc{\'o}n-Barroso}, J. 2016, \mnras, 463, 3409, \dodoi{10.1093/mnras/stw2231}

\bibitem[{{Walsh} {et~al.}(2016){Walsh}, {van den Bosch}, {Gebhardt}, {Y{\i}ld{\i}r{\i}m}, {Richstone}, {G{\"u}ltekin}, \& {Husemann}}]{walshetal2016}
{Walsh}, J.~L., {van den Bosch}, R.~C.~E., {Gebhardt}, K., {et~al.} 2016, \apj, 817, 2, \dodoi{10.3847/0004-637X/817/1/2}

\bibitem[{Ye(1998)}]{Ye1998}
Ye, J. 1998, Journal of the American Statistical Association, 93, 120, \dodoi{10.1080/01621459.1998.10474094}

\end{thebibliography}
\bibliographystyle{aasjournal}

\end{document}